\documentclass{article}

\usepackage{mathrsfs,
  fancyhdr,
  amssymb,
  amsmath,
  amsthm,
  fancybox,
  graphicx,
  epsfig,
  xtab,
  rotating,
  proof,
  color}
\newcommand{\cors}[1]{\ensuremath{\mathscr{#1}}}

\newcommand{\vdashNEAL}{\ensuremath{\vdash_{\scriptscriptstyle{\mathsf{NEAL}}}}}
\newcommand{\linear}{\ensuremath{\multimap}}
\newcommand{\lin}{\linear}
\newcommand{\lam}{\ensuremath{\lambda}} 

\newcommand{\eal}{EAL}
\newcommand{\vdashEAL}{\ensuremath{\vdash_{\mathsf{\eal}}}}

\newcommand{\contr}[4]{\ensuremath{[#1]_{#2=#3,#4}}}
\newcommand{\promote}[5]{\ensuremath{!\left(#1\right)
    \left[{}^{#2}/#3,\ldots,{}^{#4}/#5\right]}}
\newcommand{\vertarray}[2]{\ensuremath{\left\{\!\!\!\!\begin{array}{c}
        #1\\
        \vdots\\
        #2\end{array}\!\!\!\!\right\}}}

\newcommand{\FV}{\ensuremath{\mathtt{FV}}}

\newcommand{\bbbn}{\ensuremath{\mathrm{I\!N}}}
\newcommand{\Proc}{\ensuremath{\cors{P}}}
\newcommand{\Sint}{\ensuremath{\mathcal{S}}}
\newcommand{\Boxing}{\ensuremath{\mathbb{B}}}
\newcommand{\orizarray}[2]{\ensuremath{\left\{#1,\ldots,#2\right\}}}
\newcommand{\Unif}{\ensuremath{\cors{U}}}
\newcommand{\Bang}{\ensuremath{\cors{B}}}

\newtheorem{fact}{Fact}
\newtheorem{definition}{Definition}
\newtheorem{lemma}{Lemma}

\newtheorem{theorem}{Theorem}
\newtheorem{proposition}{Proposition}
\newtheorem{note}{Note}
\newtheorem{remark}{Remark}
\newtheorem{notation}{Notation}
\newtheorem{corollary}{Corollary}

\begin{document}
\title{{O}ptimizing {O}ptimal {R}eduction:\\
A Type Inference Algorithm for Elementary Affine Logic}
            
\author{Paolo Coppola\\ Universit\`a di Udine --- Dip.\ di Matematica e  Informatica
\and Simone Martini\\ Universit\`a di Bologna --- Dip.\ di Scienze dell'Informazione
}
            
            
            
            
            


\maketitle

\section*{Introduction}
\label{intro}
The optimal reduction of $\lambda$-terms (\cite{Levy:1980};
see~\cite{Asperti:Guerrini:1998} for a comprehensive account and
references) is a graph-based technique for normalization in which a
redex is never duplicated. To achieve this goal, the syntax tree of
the term is transformed into a graph, with an explicit node ({\em
  fan\/}) expressing the sharing of two common subterms (these
subterms are always variables in the initial translation of a
$\lambda$-term).  Giving correct reduction rules for these
\emph{sharing graphs} is a surprisingly difficult problem, first
solved in~\cite{Kathail:1990:phd,Lamping:1990:popl}.  One of the main
issues is to decide how to reduce two meeting fans, for which a
complex machinery and new nodes have to be added (the \emph{oracle}).
There is large class of (typed) terms, however, for which this
decision is very simple, namely those $\lambda$-terms whose sharing
graph is a proof-net of Elementary Logic, both in the
Linear~\cite{Girard:1998:i&c} (ELL) and the
Affine~\cite{Asperti:1998:lcs} (EAL) flavor.  This fact was first
observed in~\cite{Asperti:1998:lcs} and then exploited
in~\cite{Asperti:Coppola:Martini:2000:popl} to obtain a certain
complexity result on optimal reduction, where
(following~\cite{Mairson:1992:tcs}) we also showed that these
\emph{EAL-typed} $\lambda$-terms are powerful enough to encode
arbitrary computations of elementary time-bounded Turing machines. We
did not know, however, of any systematic way to derive EAL-types for
$\lambda$-terms, a crucial issue if we want to exploit in an optimal
reducer the added benefits of this class of terms. This is what we
present in this paper.

Main contribution of the paper is a type inference algorithm
(Section~\ref{sec:typeinf}), assigning EAL-types (formulas) to {\em
  type-free\/} $\lambda$-terms (more precisely: to sharing graphs
corresponding to type-free $\lambda$-terms).  We will see in
Section~\ref{sec:eal} that a typing inference for a $\lambda$-term $M$
in EAL consists of a {\em skeleton\/} -- given by the assignment of a
type to $M$ in the simple type discipline -- together with a {\em box
  assignment\/}, essential because EAL allows contraction only on
boxed terms.  The algorithm tries to introduce all possible boxes by
collecting integer linear constraints during the exploration of the
syntax tree of $M$. At the end, the integer solutions (if any) to the
constraints give specific box assignments (\emph{i.e.},
EAL-derivations) for $M$. Correctness and completeness of the
algorithm are proved with respect to a natural deduction system for
EAL, introduced in Section~\ref{sec:NEAL} together with terms
annotating the derivations.

The technique used in the paper, with minor modifications, can be used
to obtain linear logic derivations as decorations of intuitionistic
derivations, subsuming some of the results
of~\cite{Danos:Joinet:Schellinx:1995:aml,Schellinx:1994:phd}.  In this
way we may obtain linear derivations with a minimal number of boxes.
We tackle this issue in Section~\ref{sect:LL}.

A preliminary version of this work has already been
published~\cite{Coppola:Martini:2001:tlca:lncs}. Besides giving more
elaborated examples and technical details, several results are new. We
prove that all EAL types can be obtained by applying the algorithm on
the simple principal type schema; as a corollary, we may state the
decidability of the type inference problem for EAL. We show how to use
our technique to decorate full linear logic proofs. We show how the
algorithm could be extended to allow arbitrary contractions.

In~\cite{Coppola:Ronchi:2003:tlca:lncs}, the existence of a notion of
principal type schema for EAL is investigated and established.
Baillot~\cite{Baillot:2002:ifiptcs} gives a type-checking algorithm
for Light Affine Logic, but it applies only to lambda terms in normal
form. In~\cite{Baillot:2003:tr} the same author proves the
decidability of LAL type inference problem for lambda-calculus
following the approach proposed
in~\cite{Coppola:Ronchi:2003:tlca:lncs}.

\section{Elementary Affine Logic}
\label{sec:eal}
Elementary Affine Logic~\cite{Asperti:1998:lcs} is a system with
unrestricted weakening, where contraction is allowed only for modal
formulas. There is only one {\em exponential\/} rule for the modality
! ({\em of-course}, or {\em bang\/}), which is introduced at once on
both sides of the turnstile. The system is presented in
Figure~\ref{fig:EAL}, where also $\lambda$-terms are added to the
rules. We denote with $M\{N/x\}$ the usual notion of substitution of
$N$ for the free occurrences of $x$ in $M$. In the contexts (or bases)
($\Gamma$, $\Delta$, etc.)  a variable can occur only once (they are
\emph{linear}).  Observe that, according to most literature on optimal
reduction, we always write parenthesis around an application and we
assume that the scope of a $\lam$ is the minimal subterm following the
dot; as a consequence, a term like $(\lam x.M\; N)$ should be parsed
as $((\lam x.M) N)$.  Cut-elimination may be proved for EAL in a
standard way.

\begin{figure}
\begin{center}
\scalebox{.8}{
\fbox{%
\begin{tabular}{ c c c }
&&\\
$\infer[ax]{x:A\vdash x:A}{}$ & $\infer[cut]{\Gamma,\Delta\vdash M\{N/x\}:B}{
  \Gamma\vdash N:A & x:A,\Delta\vdash M:B}$&\\&&\\
$\infer[weak]{\Gamma,x:A\vdash M:B}{\Gamma\vdash M:B}$ &
$\infer[contr]{\Gamma,z:!A\vdash M\{z/x_1,z/x_2\}:B}{\Gamma,x_1:!A,x_2:!A\vdash M:B}$&\\&&\\
\hspace{2em}$\infer[\linear R]{\Gamma\vdash \lam x.M : A\linear B}{\Gamma,x:A\vdash
  M:B}$ &\hspace{2em} 
$\infer[\linear L]{\Gamma,f:A\linear B,\Delta\vdash M\{(f\  N)/x\}:C}{\Gamma\vdash N:A &
  x:B,\Delta\vdash M:C}$&\mbox{\hspace{2em}}\\&&\\
\multicolumn{3}{ c }{$\infer[!]{x_1:!A_1,\ldots,x_n:!A_n\vdash
    M:!B}{x_1:A_1,\ldots,x_n:A_n\vdash M:B}$}\\
&&\\
\end{tabular}}}
\caption{(Implicational) Elementary Affine Logic}~\label{fig:EAL}
\end{center}
\end{figure}

Given the sharing graph of a type-free $\lam$-term, we are interested
in finding a derivation of a type for it, according to
Figure~\ref{fig:EAL}. (There is a subtle point in this notion, which
is relevant for the completeness of our algorithm and which we will
discuss at the end of this section.  For the time being we may remain
informal).

A simple inspection of the rules of EAL shows that any $\lam$-term
with an EAL type has also a simple type. Indeed, the simple type (and
the corresponding derivation) is obtained by forgetting the
exponentials, which must be present in an EAL derivation because of
contraction.  Therefore, in looking for an EAL-type for a $\lam$-term
$M$, we can start from a simple type derivation for $M$ and try to
decorate this derivation (\emph{i.e.}, add !-rules) to turn it into an
EAL-derivation.  Our algorithm implements this simple idea:
 \begin{enumerate}
 \item we find all ``maximal decorations'';
 \item these decorations correspond to well formed derivations only if 
           certain linear constraints admit (integral) solutions.
 \end{enumerate}
 We informally present the main point with an example on the term
 $two\equiv\lam x y.(x (x\ y))$.  One simple type derivation
 for $two$ (expressed as a sequent derivation) is:

 \begin{displaymath}
   \infer{\scriptstyle\vdash\lam x y.(x(x\ y)):(\alpha\to\alpha)\to \alpha\to
     \alpha}{
     \infer{\scriptstyle x:\alpha\to\alpha \vdash \lam y.(x(x\ y)):\alpha\to\alpha}{ 
       \infer{\scriptstyle x:\alpha\to\alpha ,x:\alpha\to\alpha \vdash \lam y.(x(x\
         y)):\alpha\to\alpha}{
         \infer{\scriptstyle x:\alpha\to\alpha ,x:\alpha\to\alpha, y:\alpha \vdash (x(x\
           y)):\alpha}{
           \infer{\scriptstyle x:\alpha\to\alpha, y:\alpha \vdash (x\ y):\alpha}{
             \infer{\scriptstyle w:\alpha\vdash w:\alpha}{}
             &
             \infer{\scriptstyle y:\alpha\vdash y:\alpha}{}
           }
           &
           \infer{\scriptstyle z:\alpha\vdash z:\alpha}{}
         }
       }
     }
   }
 \end{displaymath}
 
 If we change every $\to$ in $\linear$, the previous derivation can be
 viewed as the skeleton of an \eal{} derivation. To obtain a full
 \eal{} derivation (if any), we need to decorate this
 skeleton with exponentials, and to check that the contraction is
 performed only on exponential formulas.
 
We first produce a \emph{maximal decoration} of the skeleton,
 interleaving $n$ !-rules after each logical rule. For
 instance
 \begin{displaymath}
 \infer{\scriptstyle x:\alpha\linear\alpha,y:\alpha\vdash (x\ y):\alpha}{
   \infer{\scriptstyle w:\alpha\vdash w:\alpha}{}
   &
   \infer{\scriptstyle y:\alpha\vdash y:\alpha}{}
 }
 \end{displaymath}
 becomes
 \begin{displaymath}
 \infer{\scriptstyle x:!^{n_2}\alpha\linear!^{n_1}\alpha,y:!^{n_2}\alpha\vdash (x\
   y):!^{n_1}\alpha}{ 
   \infer=[\scriptstyle !^{n_1}]{\scriptstyle !^{n_1}w:\alpha\vdash !^{n_1}w:\alpha}{
     \infer{\scriptstyle w:\alpha\vdash w:\alpha}{}
    }
   &
   \infer=[\scriptstyle !^{n_2}]{\scriptstyle !^{n_2}y:\alpha\vdash !^{n_2}y:\alpha}{
     \infer{\scriptstyle y:\alpha\vdash y:\alpha}{}
   }
 }
 \end{displaymath}
 where $n_1$ and $n_2$ are fresh variables. We obtain in this way a
 meta-derivation representing all \eal{} derivations with
 $n_1,n_2\in\bbbn$.
 
 Continuing to decorate the skeleton of $two$ (\emph{i.e.}, to
 interleave !-rules) we obtain
 \begin{displaymath}
     \infer{\scriptstyle x:!^{n_5+n_6}(!^{n_1+n_3}\alpha\linear
         !^{n_4} \alpha) \vdash \lam y.(x(x\
         y)):!^{n_6}(!^{n_2+n_3+n_5}\alpha\linear
         !^{n_4+n_5}\alpha)}{ 
       \infer=[\scriptstyle\!\!\!!^{n_6}]{\scriptstyle
         x:!^{n_5+n_6}(!^{n_1+n_3}\alpha\linear 
         !^{n_4} \alpha),x:!^{n_3+n_5+n_6}(!^{n_2}\alpha\linear!^{n_1}\alpha)
         \vdash \lam y.(x(x\ y)):!^{n_6}(!^{n_2+n_3+n_5}\alpha\linear
         !^{n_4+n_5}\alpha)}{
         \infer{\scriptstyle x:!^{n_5}(!^{n_1+n_3}\alpha\linear
         !^{n_4} \alpha),x:!^{n_3+n_5}(!^{n_2}\alpha\linear!^{n_1}\alpha)
         \vdash \lam y.(x(x\ y)):!^{n_2+n_3+n_5}\alpha\linear
         !^{n_4+n_5}\alpha}{
           \infer=[\scriptstyle!^{n_5}]{\scriptstyle
         x:!^{n_5}(!^{n_1+n_3}\alpha\linear 
           !^{n_4} \alpha),x:!^{n_3+n_5}(!^{n_2}\alpha\linear!^{n_1}
           \alpha), y:!^{n_2+n_3+n_5}\alpha \vdash (x(x\ y)):!^{n_4+n_5}
           \alpha}{
             \infer{\scriptstyle x:!^{n_1+n_3}\alpha\linear
             !^{n_4} \alpha,x:!^{n_3}(!^{n_2}\alpha\linear!^{n_1}
             \alpha), y:!^{n_2+n_3}\alpha \vdash (x(x\ y)):!^{n_4}
             \alpha}{
               \infer=[\scriptstyle !^{n_3}]{\scriptstyle
         x:!^{n_3}(!^{n_2}\alpha\linear!^{n_1} 
               \alpha), y:!^{n_2+n_3}\alpha \vdash (x\ y):!^{n_1+n_3}
               \alpha}{
                 \infer{\scriptstyle x:!^{n_2}\alpha\linear!^{n_1}
                   \alpha, y:!^{n_2}\alpha \vdash (x\ y):!^{n_1}
                   \alpha}{
                   \infer=[\scriptstyle!^{n_1}]{\scriptstyle
         w:!^{n_1}\alpha\vdash w:!^{n_1}\alpha}{ 
                     \infer{\scriptstyle w:\alpha\vdash w:\alpha}{}
                   }
                   &
                   \infer=[\scriptstyle!^{n_2}]{\scriptstyle
         y:!^{n_2}\alpha\vdash y:!^{n_2}\alpha}{ 
                     \infer{\scriptstyle y:\alpha\vdash y:\alpha}{}
                   }
                 }
               }
               &
               \infer=[\scriptstyle !^{n_4}]{\scriptstyle
         z:!^{n_4}\alpha\vdash z:!^{n_4}\alpha}{ 
                 \infer{\scriptstyle z:\alpha\vdash z:\alpha}{}
               }
             }
           }
         }
       }
     }
 \end{displaymath}
 The last rule---contraction---is correct in \eal{} iff the types of
 $x$ are unifiable and banged. In other words iff the following
 constraints are satisfied:
 \begin{displaymath}
     \scriptstyle n_1,n_2,n_3,n_4,n_5, n_6\in\bbbn\quad\land\quad
           \scriptstyle n_5 = n_3+n_5\quad\land\quad
          \scriptstyle n_1+n_3 = n_2\quad\land\quad
          \scriptstyle n_4=n_1\quad\land\quad
          \scriptstyle n_5+n_6\ge 1.
 \end{displaymath}
 The second, third and fourth of these constraints come from
 unification; the last one from the fact that contraction is allowed
 only on exponential formulas. These constraints are equivalent to
 \begin{displaymath}
    \scriptstyle n_1,n_5, n_6\in\bbbn\quad\land\quad
    \scriptstyle n_3 = 0\quad\land\quad
    \scriptstyle n_1 = n_2 = n_4\quad\land\quad
    \scriptstyle n_5+n_6\ge 1.
 \end{displaymath}
 Since clearly these constraints admit solutions, we conclude the
 decoration procedure obtaining
 \begin{displaymath}
   \infer{\scriptstyle\vdash\lam x y.(x(x\ y)):!^{n_5+n_6}(!^{n_1}\alpha\linear
     !^{n_1}\alpha)\linear !^{n_6}(!^{n_1+n_5}\alpha\linear
     !^{n_1+n_5}\alpha)}{ 
     \infer[]{\scriptstyle x:!^{n_5+n_6}(!^{n_1}\alpha\linear
     !^{n_1}\alpha) \vdash  
     \lam y.(x(x\ y)):!^{n_6}(!^{n_1+n_5}\alpha\linear!^{n_1+n_5}
     \alpha)}{\scriptstyle\vdots
     }
   }
 \end{displaymath}
 Thus $two$ has EAL types $!^{n_5+n_6}(!^{n_1}\alpha\linear
 !^{n_1}\alpha)\linear !^{n_6}(!^{n_1+n_5}\alpha\linear
 !^{n_1+n_5}\alpha)$, for any $n_1,n_5,n_6$ solutions of
 \begin{displaymath}
    \scriptstyle n_1,n_5, n_6\in\bbbn\quad\land\quad
    \scriptstyle n_5+n_6\ge 1.
 \end{displaymath}
 
 While simple and appealing, the technique of maximal decoration
 cannot be applied directly. The first problem is that sequent
 derivations are too constrained. There are many different (simple
 type) derivations for the same $\lam$-term, depending on the position
 of ($\linear L$) rules, contractions, cuts, etc. Given a $\lam$-term,
 we should therefore produce all possible derivations, and then
 decorate them. The problem stems from the fact that sequent
 derivations are not driven by the syntax of the term.  In fact, the
 standard simple type inference algorithm does \emph{not} use a
 sequent-style presentation, but a natural deduction one, which is
 naturally syntax-driven. This is the solution we also follow in this
 paper --- we decorate the $\lam$-term. Unfortunately, it is well
 known (see Prawitz's classical essay~\cite{Prawitz:1965}) that
 natural deduction for modal systems behave badly, since the obvious
 formulation for the modal rule (the one coinciding with rule $!$ of
 the sequent presentation) does not enjoy a substitution lemma. As a
 result, there are EAL type inferences which cannot be obtained
 directly as decoration of simple type derivations \emph{in natural
   deduction}.  Consider, for instance, the following simple type
 derivation (in the obvious natural deduction presentation of
 implicational logic) for $M = \lam x\ y\ k.(x\ y):(A\to B)\to
 A\to(C\to B)$:

\begin{displaymath}
    \infer{\vdash \lam x\ y\ k.(x\ y):(A\to B)\to A\to (C\to B)}{
      \infer{x:A\to B\vdash \lam y\ k.(x\ y):A\to (C\to B)}{
        \infer{x:A\to B, y:A\vdash \lam k.(x\ y):C\to B}{
          \infer{x:A\to B, y:A, k:C\vdash (x\ y):B}{
            x:A\to B\vdash x:A\to B
            &
            y:A, k:C\vdash y:A}
          }
        }
      }
  \end{displaymath}

It is not difficult to see that in the system of Figure~\ref{fig:EAL} there
is a derivation establishing $\vdash M: (A\lin !B)\lin A\lin!(C\lin B)$.
But no interleaving of $!$ rules into the derivation above can give this
conclusion. 

Indeed, to guarantee a substitution lemma, the modal rule for EAL in natural
deduction must be formulated:
\begin{displaymath}
\infer[box]{\Delta_1,\ldots,\Delta_n, \vdash\; !B}
           {\Delta_1\vdash !A_1 & 
             \ldots & 
             \Delta_n\vdash !A_n & 
             A_1,  \ldots, A_n \vdash B}
\end{displaymath}

This rule, given a derivation of $A_1, \ldots, A_n \vdash B$
(\emph{i.e.}, a $\lam$-term $M$ with the assignment of the type $B$
from the basis $A_1, \ldots, A_n$): (i) ``builds a box'' around $M$;
(ii) allows the substitution of arbitrary terms for the free variables
of $M$.

\begin{figure}
\begin{center}
  \scalebox{.9}{%
  \input{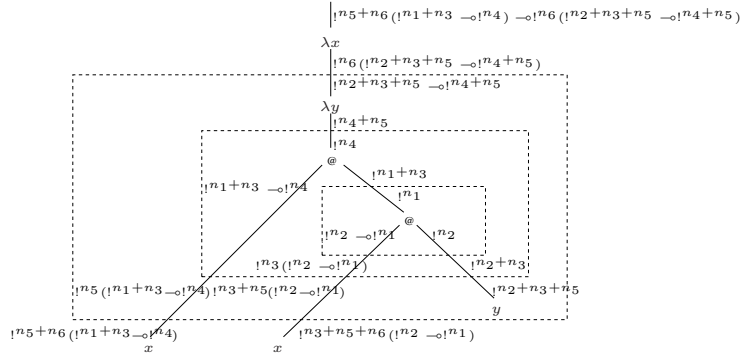}}
  \caption{Meta EAL type derivation of $two$.}~\label{fig:two_eal_typed1}
\end{center}
\end{figure}

Our algorithm will start from a simple type derivation in natural
deduction for a term $M$ (\emph{i.e.}, the syntax tree of the term
decorated with simple types) and will try to insert (all possible)
boxes around (suitable) subterms.  We will sometimes use a graphical
representation of this process. As an example,
Figure~\ref{fig:two_eal_typed1} shows the decoration of the syntax
tree of $two$ we obtained in Section~\ref{sec:eal}.

\begin{figure}
  \centering
  \scalebox{.8}{
    \input{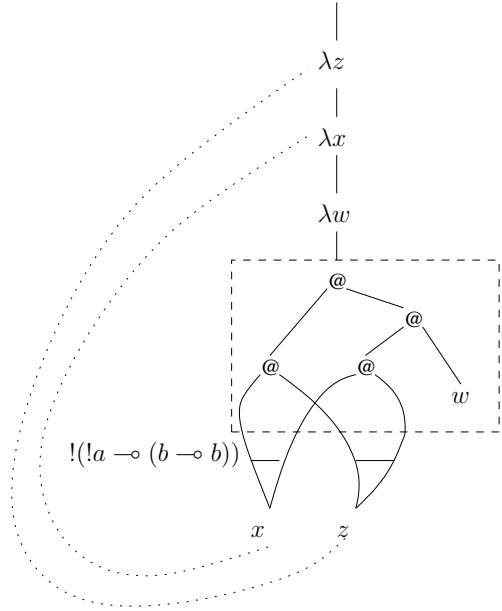}
    }
  \caption{One  decoration of $\lam z\ x\ w.((x\ z)\ ((x\ z)\ w))$:
  the fan faces a lambda.} 
   \label{fig:OKcontr}
\end{figure}

\begin{figure}
  \centering
  \scalebox{.8}{
    \input{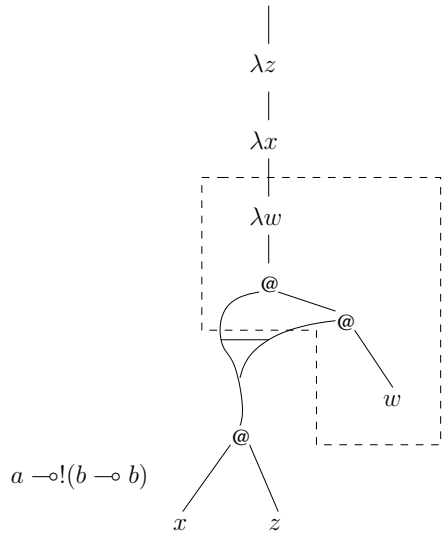}
    }
    \caption{Another decoration of  $\lam z\ x\ w.((x\ z)\ ((x\ z)\ w))$:
      the fan faces an application.}
  \label{fig:only-typeable-with-contraction}
\end{figure}

We are finally in the position to introduce formally the notion of
EAL-typing for $\lambda$-terms.  Recall that our main goal is to
mechanically check whether a pure $\lam$-term could be optimally
reduced without the need of the oracle. While we lack a general
characterization of this class of terms, we know that it contains any
sharing graph coding the skeleton of a sequent proof in EAL.  We
already observed, however, that a single $\lam$-term may correspond to
more than one (sequent or natural deduction) proof. The position of
the contraction is especially relevant in this context. Indeed,
consider the term $M=\lam z\ x\ w.((x\ z)\ (x\ z)\ w)$.  Among the
(infinite) EAL sequent derivations having $M$ as a skeleton consider
the following two fragments:
\begin{equation}
\label{eq:OKcontr}
\infer[\lin R]{z:!a,x:!(a\lin (b\lin b))\vdash \lam w.((x\ z)\ ((x\ z)\
  w)):(!b\lin !b)}{ 
  \infer=[contr,contr]{ z:!a,x:!(a\lin (b\lin b)),w:!b \vdash
    ((x\ z)\ ((x\ z)\ w)) : !b}{ 
    \infer[!]{z_1:!a,z_2:!a,x_1:!(a\lin (b\lin
      b)),x_2:!(a\lin (b\lin b)),w:!b \vdash ((x_1\ z_1)\ ((x_2\
      z_2)\ w)) : !b}{ 
      \infer{z_1:a,z_2:a,x_1:a\lin (b\lin
        b),x_2:a\lin (b\lin b),w:b \vdash ((x_1\ z_1)\ ((x_2\ z_2)\
        w)) : b}{ 
        \vdots
      }
    }
  }
}
\end{equation}
and
\begin{equation}
\label{eq:KOcontr}
  \infer[\lin L]{z:a, x:a\lin!(b\lin b)\vdash \lam w.((x\ z)\ ((x\ z)\ w)):
    !(b\lin b)}{
    z:a\vdash z:a
    &
    \infer[contr]{k:!(b\lin b)\vdash \lam w.(k\ (k\ w)):!(b\lin b)}{
      \infer[!]{k_1,k_2:!(b\lin b)\vdash \lam w.(k_1\ (k_2\ w)):!(b\lin
        b)}{
        \infer{k_1,k_2:b\lin b\vdash \lam w.(k_1\ (k_2\ w)):b\lin b)}{
          \vdots}
      }
    }
  }
\end{equation}
If we display these derivations as annotated syntax tree with explicit
fan nodes for contraction (that is, as sharing graphs), we obtain
Figure~\ref{fig:OKcontr} for the derivation~(\ref{eq:OKcontr}), and
Figure~\ref{fig:only-typeable-with-contraction}
for~(\ref{eq:KOcontr}).


Both graphs are legal EAL sharing graphs, but only the first is a
possible initial translation of $M$ as a sharing graph, since in
initial translations the fan nodes are used to share (contract) only
variables, before abstracting them.  Although our technique could be
extended to cope with arbitrary contractions (see
Section~\ref{sect:conclusions}), we present it as a type inference
algorithm for initial translations of type-free $\lam$-terms,
according to our original aim to use it as a tool in an optimal
reducer. This is the motivation for the following notion.
\begin{definition}
  \label{def:officialEALtyping}
  A type-free $\lam$-term $M$ has EAL type $A$ from the basis $\Gamma$
  (write: $\Gamma \vdashEAL M:A$) iff there is a derivation of
  $\Gamma\vdash M:A$ in the system of Figure~\ref{fig:EAL} whose
  corresponding sharing graph does not have any fan node facing an
  application node.
\end{definition}

\begin{remark}
  It is possible to formulate the previous definition directly in
  terms of sequent derivations, without any reference to the notion of
  sharing graph.  It could be proved that $\Gamma \vdashEAL M:A$ iff
  there is a sequent derivation of $\Gamma\vdash M:A$ where all
  contractions either are immediately followed by $\lin R$, or are at
  the end of the derivation. However, the ``only if'' part is not
  trivial.  In going from a sequent derivation to a sharing graph, in
  fact, we loose any information regarding the position of cuts and
  (to some extent) of $\lin L$. Therefore, given a term $M$ for which
  $\Gamma \vdashEAL M:A$ (that is, given a sharing graph that could be
  decorated with EAL-types and boxes) there are many sequent
  derivations corresponding to the skeleton coded by this sharing
  graph.  Not all these derivations satisfy the constraint expressed
  by the ``only if'' part. It can be shown, however, that among these
  derivations there is one in which the constraint is satisfied.  This
  could be obtained by using the notion of canonical form of an EAL
  derivation, introduced and exploited
  in~\cite{Coppola:Ronchi:2003:tlca:lncs}.
\end{remark}

\begin{remark}                            
  There exist simply typeable terms without any \eal{} type. For
  instance the $\lam$-term
\begin{displaymath}
  (\lam  n.(n\ \lam y.(n\  \lam z.y))\  \lam x.(x\  (x\ y)))
\end{displaymath}
has a simple type, but  no \eal{} decoration (see Appendix~\ref{ex:non-typeable} for an
analysis).
\end{remark}

\section{Type inference}
\label{sec:typeinf}
The type inference algorithm is given as a set of inference rules,
specifying several functions. The complete set of rules is given in
Section~\ref{section:type-inf-rules}; the properties of the algorithm
will be stated and proved in Section~\ref{sect:properties}.  We start
in the next section with the detailed discussion of an example, which
will also introduce the various rules and the problems they have to
face.

 \subsection{Example of type inference}
 \label{subsec:example}
 A class of types for an \eal-typeable term can be seen as a
 decoration of a simple type with a suitable number of boxes.
 \begin{definition}
 \label{def:generalEALtype}
 A \emph{general \eal-type} $\Theta$ is generated by the following
 grammar:
 \begin{displaymath}
 \Theta ::= !^{n_1+\cdots+n_k}o | !^{n_1+\cdots+n_k}(\Theta
 \linear \Theta)
 \end{displaymath}
 where $k\ge 0$ and $n_1,\ldots,n_k$ are variables ranging on $\bbbn$. 
 \end{definition}
 We shall illustrate our algorithm on the term $(\lam n.\lam y.((n\ 
 \lam z.z)\ y)\ \lam x.(x\ (x\ \lam w.w))):o \to o$, whose
 \emph{simple} type derivation in natural deduction is given in
 Figure~\ref{fig:simple_type_derivation} ($I_\alpha$ stands for
 $\alpha\to \alpha$).

 \begin{figure}
   \scalebox{.7}{\ensuremath{ \infer{\vdash (\lam n.\lam y.((n\ \lam
         z.z)\ y)\ \lam x.(x\ (x\ \lam w.w))):o \to o}{ \infer{\vdash
           \lam n.\lam y.((n\ \lam z.z)\ y):(I_{I_o}\to I_o)\to I_o}{
           \infer{n:I_{I_o}\to I_o\vdash \lam y.((n\ \lam z.z)\ 
             y):I_o}{ \infer{n:I_{I_o}\to I_o, y:o\vdash ((n\ \lam
               z.z)\ y):o}{ \infer{n:I_{I_o}\to I_o\vdash(n\ \lam
                 z.z):I_o}{ n:I_{I_o}\to I_o\vdash n:I_{I_o}\to I_o &
                 \infer{\vdash \lam z.z:I_{I_o}}{ z:I_o\vdash z:I_o }
                 } & y:o\vdash y:o } } } & \infer{\vdash \lam x.(x\ 
           (x\ \lam w.w)):I_{I_o}\to I_o}{ \infer{x:I_{I_o}\vdash (x\ 
             (x\ \lam w.w)):I_o}{ x:I_{I_o}\vdash x:I_{I_o} &
             \infer{x:I_{I_o}\vdash (x\ \lam w.w):I_o}{
               x:I_{I_o}\vdash x:I_{I_o} & \infer{\vdash \lam
                 w.w:I_o}{w:o\vdash w:o} } } } } }}
 \caption{Simple type derivation of $(\lam n.\lam y.((n\ \lam z.z)\ y)\
   \lam x.(x\ (x\ \lam w.w))):o \to
   o$}\label{fig:simple_type_derivation}
 \end{figure}
 
 The algorithm searches for the leftmost innermost subterm for which 
there is no assignment of an EAL-type yet. 
 In this case, it is the variable
 \begin{displaymath}
 n:\left(\left((o\to o)\to(o\to
   o)\right)\to (o\to o)\right).
 \end{displaymath}
 Its most general \eal-type is obtained from its simple type by adding
 $p_i$ modalities wherever possible. This is the r\^ole of the
 function $\Proc$:
 \begin{gather}
 \label{eq:processing1}
 \frac{}{\Proc(o)=!^p o}{}
\end{gather}
\begin{gather}
 \label{eq:processing2}
 \frac{ \Proc(\sigma)=\Theta \hfill \Proc(\tau)=\Gamma }{
   \Proc(\sigma\to\tau)=!^p (\Theta \linear \Gamma) }.
 \end{gather}
 The main function of the algorithm---the type synthesis function
 $\Sint$---may now be applied. In the case of a variable $x$ of simple
 type $\sigma$ the rule is:
   \begin{equation}\label{eq:syntvariable}
     \infer{\Sint(x:\sigma)= \langle  \Theta , \{x:\Theta\}, \emptyset,
       \emptyset\rangle }{ \Proc(\sigma)=\Theta }
   \end{equation}
   Observe that, given a term $M$ of simple type $\sigma$,
   $\Sint(M:\sigma)$ returns a quadruple:
   \begin{center}
     $\langle$general \eal-type, base\footnote{A base here is a
       \emph{multiset} where multiple copies of $x:\Theta$ may be
       present.}  $\{x_i:\Theta_i\}_i$ of pairs (variable:general
     \eal-type), set of linear constraints, critical
     points\footnote{We will discuss critical points in a
       moment.}$\rangle$.
   \end{center}
   In our example we obtain:
   \begin{equation}
     n:!^{p_1}\left(!^{p_2}\left(!^{p_3}    (!^{p_4}o\linear    !^{p_5}
         o)\linear    !^{p_6}(!^{p_7}o\linear   !^{p_8}o)\right)\linear
       !^{p_9}(!^{p_{10}}o\linear !^{p_{11}}o)\right)
   \end{equation}
   for any $p_i\in \bbbn, 1\leq i\leq 11$. In the following we will
   not explicit the ``$\in \bbbn$'' for any variable we will
   introduce, being this constraint implicated by
   Definition~\ref{def:generalEALtype}.
 \begin{notation}
   We will write $(n\linear m)$ instead of $(!^n o\linear
   !^m o)$, for a better reading.
 \end{notation}
 
 Analogously, $z:(o\to o)$ is typed
 \begin{equation}
   z: p_{12}(p_{13}\linear p_{14})
 \end{equation}
 It is now the turn of the subterm $\lam z.z$. The type synthesis rule
 for an abstraction $\lam x.M$, where $x$ occurs in $M$, takes the
 following steps:
 \begin{enumerate}
 \item infer the \eal-type for $M$;
 \item add \emph{all possible boxes} around $M$ (function $\Boxing$,
   which will be described later); the algorithm tries to build all
   possible decorations\footnote{More precisely it builds all possible
     decorations without exponential cuts and with some other
     properties listed in Theorem~\ref{theor:completeness}.
     Decorations of these kinds are sufficient for the completeness of
     the algorithm.}  that in the case of an abstraction $\lam x.M$
   are the decorations of all subterms of $M$, already build by
   inductive hypothesis, plus all possible box-decorations of the
   whole $M$, performed at this stage of the inference by function
   $\Boxing$, plus all possible box decorations of $\lam x.M$,
   eventually performed at the next step of the inference procedure;
 \item contract all the types of abstracted variable $x$ (function
   $\mathcal{C}$, which will be described later).
 \end{enumerate}
 The rule is the following:
 \begin{equation}\label{eq:synt_lamx}
   \frac{
     \begin{array}{l}
       \mathcal{C}(\Theta_1,\ldots,\Theta_h)=A_3\\
       \Boxing\left(M, B_1, \Gamma_1, cpts\cup
       \vertarray{sl_1(x)}{sl_k(x)}, A_1\right)=\left\langle B\cup
       \vertarray{x:\Theta_1}{x:\Theta_h}, \Gamma, A_2 \right\rangle\\
       \Sint(M:\tau)=\left\langle
       \Gamma_1,B_1, A_1,cpts\cup \orizarray{sl_1(x)}{sl_k(x)}
 \right\rangle\\
     \end{array}
     }{\Sint(\lam x.M:\sigma\to\tau)=\left\langle
       \Theta_1\linear\Gamma, B, \left\{\begin{array}{c}A_2\\
           A_3\end{array}\right., cpts \right\rangle}
 \end{equation}
 In our example, there is only one occurrence of $z$ and therefore the
 contraction function $\mathcal{C}$ is called with only one type and
 does not produce any constraint.  Also the boxing function $\Boxing$
 produce no result, being called on a variable, \emph{i.e.}, it acts as the
 identity returning a triple with the same base, type and (empty, in
 this case) set of constraints:
 \begin{multline*}
   \Boxing(z,\{z:{p_{12}}({p_{13}}o\linear
   {p_{14}}o)\},{p_{12}}({p_{13}}o\linear
   {p_{14}}o),\emptyset,\emptyset) =\\ \langle
   \{z:{p_{12}}({p_{13}}o\linear
   {p_{14}}o)\},{p_{12}}({p_{13}}o\linear {p_{14}}o),\emptyset\rangle.
 \end{multline*}
 The r\^ole of $cpts$ and $sl$ will be discussed in the context of the
 critical points, below.  Coming back to our example, for $\lam
 z.z:((o\to o)\to(o\to o))$ we infer the \eal-type
 \begin{equation}
   \lam  z.z: p_{12}(p_{13}\linear  p_{14})\linear p_{12}(p_{13}\linear
   p_{14})
 \end{equation}
 When the algorithm infers the \eal-type for $(n\ \lam z.z):(o\to o)$,
 it:
 \begin{enumerate}
 \item adds all possible boxes around the argument $\lam z.z$ with the
   boxing function, that in this case adds $b_1$ boxes around $\lam
   z.z$ returning a triple with the same base, $b_1$ banged type and
   unmodified set of (again empty) constraints:
   \begin{multline*}
     \Boxing(\lam z.z,\emptyset,p_{12}(p_{13}\linear p_{14})\linear
     p_{12}(p_{13}\linear p_{14}),\emptyset,\emptyset)=\\
     =\langle \emptyset, b_1(p_{12}(p_{13}\linear p_{14})\linear
     p_{12}(p_{13}\linear p_{14})), \emptyset\rangle
   \end{multline*}
 \item imposes the \eal-type of $n$ to be functional, \emph{i.e.},  the
   constraint
     \begin{equation}\label{eq:first_constraint}
       \fbox{\ensuremath{p_1 = 0}}
     \end{equation}
   \item unifies the \eal-type of the boxed $\lam z.z$ with the
     argument part of the \eal-type of $n$:
     \begin{displaymath}
       \Unif\left(\begin{array}{c}
           b_1(p_{12}(p_{13}\linear p_{14})\linear p_{12}(p_{13}\linear
           p_{14})),\\
           p_2(p_3 (p_4 \linear p_5)\linear p_6(p_7\linear p_8))
           \end{array}\right).
     \end{displaymath}
     Observe that the implicational structure of the types is already
     correct, since we start from a simple type derivation. Therefore,
     unification only produces a set of constraints on the variables
     used to indicate boxes. In our example, we get the constraints:
     \begin{equation}
       \left\{\begin{array}{ll}
           b_1    & = p_2\\
           p_{12} & = p_3\\
           p_{13} & = p_4\\
           p_{14} & = p_5\\
           p_{12} & = p_6\\
           p_{13} & = p_7\\
           p_{14} & = p_8
         \end{array}\right.\quad\Leftrightarrow\quad
       \fbox{\ensuremath{\left\{\begin{array}{lll}
               b_1 & = p_2 &\\
               p_3 & = p_6 & = p_{12}\\
               p_4 & = p_7 & = p_{13}\\
               p_5 & = p_8 & = p_{14.}
         \end{array}\right.}}
     \end{equation}
 \end{enumerate}
     \begin{figure}
       \begin{center}
         \scalebox{.9}{%
         \input{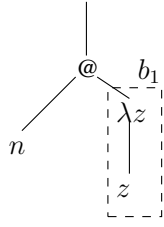}}
         \caption{Decoration of $(n\ \lam z.z)$.}
         \label{fig:ex01}
       \end{center}
     \end{figure}
     The type synthesis rule\footnote{We will explain $\doublecup$
       later.} for an application, provided that $M$ and $N$ are not
     applications themselves, is:
 \begin{equation}\label{eq:synt_app}
   \frac{
   \begin{array}{l}
     \Unif(\Theta_1,\Theta_3)=A_4\\
     \Boxing(N,B_2,\Theta_2,cpts_2,A_2)=\langle
     B_3,\Theta_3,A_3\rangle\\
     \Sint(N:\sigma)=\langle \Theta_2,B_2,A_2,cpts_2\rangle\\
     \Sint(M:\sigma\to\tau)=\langle!^{\sum n_{i}}(\Theta_1\linear
     \Gamma), B_1,A_1,cpts_1\rangle
   \end{array}}{
   \Sint((M\     N):\tau)=\left\langle     \Gamma,     B_1\cup     B_3,
     \left\{\begin{array}{c}                         A_1\\A_3\\A_4\\\sum
         n_{i}=0\end{array}\right.,                     cpts_1\doublecup
     cpts_2\right\rangle }
 \end{equation}
 Figure~\ref{fig:ex01} shows the decoration obtained so far:
 \begin{equation}
   n: b_1(p_3  (p_4 \linear  p_5)\linear p_3 (p_4  \linear p_5))\linear
   p_9(p_{10}\linear  p_{11})  \vdash  (n\ \lam  z.z):p_9(p_{10}\linear
   p_{11}).
 \end{equation}
 
 Next step is the inference of a general \eal-type $p_{15}$ for $y:o$.
 Then the algorithm starts to process $((n\ \lam z.z)\ y):o$.  As
 before, the algorithm
 \begin{enumerate}
 \item applies $\Boxing$ to the argument $y$ (a void operation here,
   since the boxing function does nothing for variables);
 \item imposes the \eal-type of $(n\ \lam z.z)$ to be functional:
   \begin{equation}\label{eq:constr_mod}
     p_9 =0.
   \end{equation}
 \item unifies the \eal-types, to make type-correct the application:
   \begin{equation}
     \Unif(p_{10},p_{15})=\quad \fbox{\ensuremath{p_{10}=p_{15.}}}
   \end{equation}
   However, the present case is more delicate than the application we
   treated before, since the function part is already an application.
   Two consecutive applications in $((n\ \lam z.z)\ y)$ indicates that
   more than one decoration is possible. Indeed, there can be several
   derivations building the same term, that can be
   differently decorated.  The issue is better appreciated if we look
   ahead for a moment and we consider the term $\lam y.((n\ \lam z.z)\ 
   y)$.  There are two (simple) sequent derivations for this term,
   both starting with the term $(x\ y):o$, for $x:o\to o, y:o$. The
   first derivation, via a left $\to$-rule, obtains $((n\ \lam z.z)\ 
   y) : o$; then it bounds $y$, giving $\lam y.((n\ \lam z.z)\ y) : o
   \to (o\to o)$.  The second derivation permutes the rules: it starts
   by binding $y$, obtaining $\lam y.(x\ y)$ and only at this point
   substitutes $(n\ \lam z.z)$ for $x$, via the left $\to$-rule. When
   we add boxes to the two derivations, we see this is a
   \emph{critical} situation.  Indeed, in the first derivation we may
   box $(x\ y)$, then $((n\ \lam z.z)\ y)$ and finally $\lam y.((n\ 
   \lam z.z)\ y)$.  In the second, we box $(x\ y)$, then $\lam y.(x\ 
   y)$ and finally the whole term.  The two (incompatible) decorations
   are depicted in the two bottom trees of
   Figure~\ref{fig:superimposed_ex}.  The critical edge---where the
   boxing radically differs---is the root of the subtree for $((n\ 
   \lam z.z)\ y)$, corresponding to the $x$ that is substituted for in
   the left $\to$-rule.  Let us then resume the discussion of the type
   inference for this term. At this stage we collect the
   \emph{critical point}, marked with a star in Figure~\ref{fig:ex02},
   indicating the presence of two possible derivations.
 \begin{figure}
    \begin{center}
      \scalebox{.7}{%
      \input{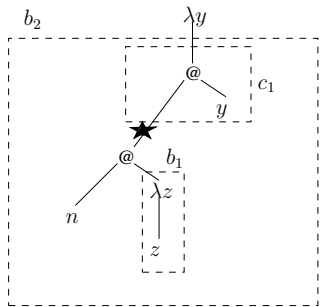}}
      \caption{Critical point in the decoration of $\lam y.((n\ \lam
        z.z)y)$.} 
       \label{fig:ex02}
    \end{center}
 \end{figure}
 When, in the future, it will be possible to add boxes, for example
 $b_2$ in Figure~\ref{fig:ex02} during the type inference of $\lam
 y.((n\ \lam z.z)\ y)$, the algorithm will consider the critical point
 as one of the closing points of such boxes, $c_1$ in
 Figure~\ref{fig:ex02}, eventually modifying the constraint in
 Equation~\eqref{eq:constr_mod} that impose type of $(n\ \lam z.z)$ to
 be functional and not exponential.  Indeed, for completeness, the
 algorithm must take into account all possible derivations. When there
 will be more than one critical point, at every stage of the type
 inference, when it is possible to apply a !  rule, the algorithm will
 compute all possible combinations of the critical points (see
 Figure~\ref{fig:critical_points}, showing a schematic example with
 two critical points) eventually modifying some constraints.  We call
 \emph{slices}\footnote{We thank Philippe Dague for useful discussions
   and suggestions on the calculation of critical points.} such
 combinations of critical points; they are the data maintained by the
 algorithm and indicated in the rules as $cpts$. The task of combining
 the two lists of slices collected during the type inference of the
 function and argument part of an application is performed by
 $\doublecup$, whose rules are given in
 Section~\ref{appendix:subsec:productunion}.
 \begin{figure}
     \begin{center}
       \includegraphics[scale=0.45]{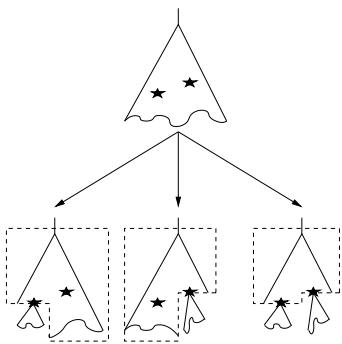}
       \caption{Combinations of two critical points.}
       \label{fig:critical_points}
     \end{center}
 \end{figure}
 \begin{definition}
   The list of \emph{free variable occurrences} of a lambda term $M$
   is defined in the following way:
 \begin{enumerate}
 \item $\mathtt{FVO}(x)=[x]$;
 \item $\mathtt{FVO}(\lam x.M)=\mathtt{FVO}(M)-x$;
 \item $\mathtt{FVO}((M_1\ M_2))=\mathtt{FVO}(M_1)::\mathtt{FVO}(M_2)$
   (the concatenation of lists).
 \end{enumerate}
 \end{definition}
   \begin{definition}
   \label{def:slice}
     A \emph{slice} is a set of pairs (constraint, list of free
     variable occurrences) as in the following\footnote{$A^j$ means
       the $j$-th row of the matrix $A$, \emph{i.e.}, the $j$-th constraint.}:
     \begin{displaymath}
     sl=\left\{(A^{j_1},[y_{1_1},\ldots,y_{1_h}]),\ldots,
       (A^{j_k},[y_{k_1},\ldots,y_{k_h}])\right\}
     \end{displaymath}
     A slice corresponds to a combination of critical points.
 \end{definition}
 In our example the algorithm collects the slice $(p_9=0,\lbrack
 n\rbrack)$. Notice that a slice partitions the set of free variable
 occurrences in a derivation: it marks the set of variable occurrences
 whose types should not be modified when the box is added. This is the
 intuitive meaning of the set of free variable occurrences in the data
 structure we use.

 \begin{notation}
\label{notation:slice}
 \begin{itemize}
 \item $sl(x)$ indicates a slice having $x$ as an element of every
   list of variables in it.
 \item $x\in sl$ if and only if there exists one element of $sl$ whose
   list of variables contains $x$.
 \item $A^j\in sl$ if and only if there exists one element of $sl$
   whose constraint is $A^j$.
 \item Being $A^j$ the constraint $\pm n_{j_1}\pm\cdots\pm n_{j_k} =
   0$, $A^j-n$ corresponds to the constraint $\pm n_{j_1}\pm\cdots\pm
   n_{j_k} -n = 0$.
 \end{itemize}
 \end{notation}
 The general type inference rule for the application we are
 considering now, \emph{i.e.}, $((M_1\ M_2)\ N)$ when $N$ is not an
 application, is the following:
 \begin{equation}\label{eq:synt_appappm}
   \frac{
   \begin{array}{l}
     cpts = \left(cpts_1\cup \{(\sum n_{i} = 0,\mathtt{FVO}((M_1\
       M_2)))\}\right)\doublecup cpts_2\\
     \Unif(\Theta_1,\Theta_3)= A_4\\
     \Boxing(N,B_2,\Theta_2,cpts_2,A_2)=\langle
     B_3,\Theta_3,A_3\rangle\\
     \Sint(N:\sigma)=\langle \Theta_2, B_2, A_2, cpts_2 \rangle\\
     \Sint((M_1\ M_2):\sigma\to\tau)=\langle !^{\sum
       n_{i}}(\Theta_1\linear\Gamma), B_1, A_1, cpts_1\rangle
   \end{array}
   }{\Sint((M_1\   M_2)\   N):\tau)   =   \left\langle   \Gamma,B_1\cup
     B_3,\left\{\begin{array}{c}A_1\\A_3\\A_4\\\sum                n_{i}
         =0\end{array}\right., cpts\right \rangle}
 \end{equation}
 In the example case we obtain:
 \begin{equation}\left\{
 \begin{array}{rcl}
   n & :& b_1(p_3 (p_4 \linear p_5)\linear p_3 (p_4 \linear
   p_5))\\&&\hspace{8em}\linear p_9(p_{10}\linear p_{11}),\\
 y & :& p_{10}\end{array}\right\} \vdash  ((n\ \lam z.z)\ y):p_{11}
 \end{equation}
 and critical points $cpts = \{(p_9=0,\lbrack n\rbrack)\}$.
 \end{enumerate}
 
 Typing $\lam y.((n\ \lam z.z)\ y):o\to o$ involves
 rule~\eqref{eq:synt_lamx}, the same we used for $\lam z.z$, but now the
 boxing procedure $\Boxing$ is called on a subterm that is not a
 single variable. The complete set of rules for $\Boxing$ is the
 following:
 \begin{equation}                                
 \label{eq:box1}%
 \infer{\Boxing(x,B,\Gamma,cpts,A)=\langle B,\Gamma,A\rangle}{}
 \end{equation}
 Boxing of a variable produces no changes in the base, type and set of
 constraints. 
 \begin{equation}                                
 \label{eq:box2}%
 \infer{\Boxing(M,B,\Gamma,cpts,A)=\langle !^b B_1, !^b \Gamma_1, A_1
   \rangle}{\Bang(B,\Gamma,cpts,A)=\langle B_1, \Gamma_1, A_1 \rangle}
 \end{equation}
 $\Bang$ takes care of the list of critical points, by adding boxes
 ``inside'' the term as in Figure~\ref{fig:critical_points}; at the
 end, $\Boxing$ adds $b$ boxes ``around'' the term.
 \begin{equation}
 \label{eq:box_critical1}%
 \infer{\Bang(B,\Gamma,\emptyset,A)=\langle
   B,\Gamma,A\rangle}{}
 \end{equation}
 \Bang{} with no critical points produces no changes.
 \begin{equation}
 \label{eq:box_critical2}%
 \infer{\Bang(\{x_i:\Theta_i\}_i,\Gamma,\{sl\}\cup cpts,A)=\langle
   B,\Delta,A_1\rangle}{
   \begin{array}{l}
   \Bang\left(B_1,!^c \Gamma,cpts,A_2\right) = \langle B,\Delta,A_1\rangle\\
   B_1=\left\{x_i:\left\{\begin{array}{ll}
         !^c\Theta_i & \qquad x_i\notin sl\\
         \Theta_i & \qquad x_i\in sl
         \end{array}\right.\right\}_i
   \\
   A_2 = \left(\left\{\begin{array}{ll}
         A^j &\qquad A^j\notin sl\\
         A^j-c &\qquad A^j \in sl
         \end{array}\right.\right)_j
   \end{array}
   }
 \end{equation}
 
 Therefore, rule~\eqref{eq:synt_lamx} gives in our case:
 \begin{multline}
   \Sint(\lam y.((n\ \lam z.z)\ y):o\to o) =\\
   \left\langle\begin{array}{c}
       b_2+c_1+p_{10}\linear b_2+c_1+p_{11},\\
       \left\{n:b_2(b_1(p_3(p_4\linear p_5)\linear p_3(p_4\linear
         p_5)) \linear  p_9(p_{10}\linear p_{11}))\right\},\\
       \left\{\begin{array}{c}\vdots\\p_9 - c_1 = 0\\ \vdots
   \end{array} \right.,\\ \{(p_9-c_1=0,[n])\}
     \end{array}\right\rangle
 \end{multline}
 where $p_9-c_1 = 0$ is the unique constraint
 (Equation~\eqref{eq:constr_mod}) modified by $\Boxing$. The
 decoration obtained is shown in Figure~\ref{fig:ex02}. Observe that,
 at this stage, the presence of incompatible derivations does not show
 up yet. It will be taken into account as soon as we will try to box a
 superterm of the one we just processed. If $\lam y.((n\ \lam z.z)\ 
 y)$ would be the whole term, on the contrary, an additional call to
 the function $\Boxing$ would be performed, see the
 rule~\eqref{eq:synt} for function $\cors{S}$.
 
 When the algorithm processes $\lam n.\lam y.((n\ \lam z.z)\ 
 y):(((o\to o)\to (o\to o))\to (o\to o))\to (o\to o)$ it applies again
 rule~\eqref{eq:synt_lamx}.  It adds $c_2$ boxes passing through the
 critical point and $b_3$ boxes around the term, obtaining:
 \begin{multline}
   \Sint(\lam n.\lam y.((n\ \lam z.z)\ y):(((o\to o)\to (o\to o))\to
   (o\to o))\to (o\to o)) =\\
   \left\langle\begin{array}{c}
       \begin{array}{r}
         b_3+b_2(b_1(p_3(p_4\linear p_5)\linear p_3(p_4\linear
         p_5))\linear p_9(p_{10}\linear p_{11}))\hspace*{2em}\\
         \linear
         b_3+c_2(b_2+c_1+p_{10}\linear b_2+c_1+p_{11})
         \end{array},\\
         \emptyset,\\
         \left\{\begin{array}{c}
             \vdots\\
             \fbox{\ensuremath{p_9-c_1-c_2 =0}}\\
             \vdots
           \end{array}\right.,\\
         \emptyset
     \end{array}\right\rangle
 \end{multline}
 where $p_9-c_1-c_2 =0$ is the unique constraints modified at this
 stage of the type synthesis.
 
 The critical point $(p_9-c_2-c_2=0,\lbrack n\rbrack)$ is removed. In
 fact, to bound $n$, the substitution of $n (\lam z.z)$ for $x$ has to
 be already performed. It does not make sense to derive first $\lam
 n.\lam y.(x\ y)$, add boxes, and then substitute $n (\lam z.z)$ for
 $x$, since this would be a free-variable catching substitution.
 \begin{figure}
   \begin{center}
     \scalebox{.6}{%
     \input{lam_n.lam_y.n_lam_z.z_yealtyped.pstex_t}}
     \caption{}
     \label{fig:ex03}
   \end{center}
 \end{figure}
 
 Figure~\ref{fig:ex03} shows the decoration obtained. Notice that
 boxes $c_2$ and $b_2$ belong to the two \emph{incompatible
   \eal-derivations} we already discussed before.  The algorithm
 maintains at the same time these derivations guaranteeing (see
 Lemma~\ref{lemma:superposition}) that if the final solution
 instantiates two incompatible derivations, we can always calculate an
 equivalent \eal-derivation (Figure~\ref{fig:superimposed_ex} shows
 the two possible derivations for our example).
 \begin{figure}
   \begin{center}
     \scalebox{.4}{ \input{superimposed_boxes.pstex_t}}
     \caption{Superimposed derivations.}
     \label{fig:superimposed_ex}
   \end{center}
 \end{figure}
 
 Going on with the type synthesis, the algorithm starts processing the
 leftmost occurrence of $x$ in $(x\ (x\ \lam w.w))$.  We use
 superscripts $(1)$ and $(2)$ to discriminate the right and left
 occurrence, respectively.  For the leftmost---$x^{(2)}:(o\to o)\to
 (o\to o)$---we infer the \eal-type
 \begin{equation}
   p_{16}(p_{17}(p_{18}\linear    p_{19})\linear   p_{20}(p_{21}\linear
   p_{22}));
 \end{equation}
 analogously, for the rightmost $x^{(1)}:(o\to o)\to(o\to o)$ we get
 the \eal-type
 \begin{equation}
   p_{23}(p_{24}(p_{25}\linear    p_{26})\linear   p_{27}(p_{28}\linear
   p_{29})).
 \end{equation}
 The \eal-type of $w:o$ is $p_{30}$ and then $\lam w.w:o\to o$ is
 typeable in \eal{} with type $p_{30}\lin p_{30}$.
 
 The innermost application $(x^{(1)}\ \lam w.w)$ is typed
 $p_{27}(p_{28}\linear p_{29})$, once we have imposed
 \begin{equation}
   \fbox{\ensuremath{p_{23}=0},}
 \end{equation}
 we have boxed $\lam w.w$ with $b_4$ boxes, and we have unified the
 types
 \begin{equation}
   \Unif(p_{24}(p_{25}\linear  p_{26}),b_4(p_{30}\lin  p_{30}))   =  
   \fbox{\ensuremath{\left\{\begin{array}{l}
           b_4 = p_{24}\\
           p_{25} = p_{26} = p_{30}.
   \end{array}\right.}}
 \end{equation}
 When the algorithm processes $(x^{(2)}\ (x^{(1)}\ \lam w.w))$, it
 adds $b_5$ boxes around the argument, imposes
 \begin{equation}
   \fbox{\ensuremath{p_{16}=0}}
 \end{equation}
 and unifies the types
 \begin{equation}\label{eq:mod_constr2}
   \Unif(p_{17}(p_{18}\linear  p_{19}),b_5+p_{27}(p_{28}\lin p_{29})) 
   =  \left\{\begin{array}{l}
       p_{17} = b_5 + p_{27}\\
       \fbox{\ensuremath{\begin{array}{l}
             p_{18} = p_{28}\\
             p_{19} = p_{29}.
   \end{array}}}
   \end{array}\right.
 \end{equation}
 Moreover, the presence of two consecutive applications makes the
 algorithm collect a new critical point $(p_{17} = b_5 +
 p_{27},\lbrack x^{(1)}\rbrack)$. The derivation obtained is:
 \begin{equation}
   \left\{\begin{array}{c}
       x^{(1)}:b_5(b_4(p_{25}\lin p_{25})\lin p_{27}(p_{18}\lin p_{19})),\\
       x^{(2)}:p_{17}(p_{18}\lin p_{19})\lin p_{20}(p_{21}\lin p_{22})
 \end{array}\right\}
 \vdash (x^{(2)}\ (x^{(1)}\ \lam w.w)):p_{20}(p_{21}\lin p_{22})
 \end{equation}
 and its decoration is shown in Figure~\ref{fig:ex04}.
 \begin{figure}
   \begin{center}
     \scalebox{.7}{%
     \input{x_x_lam_w.wealtyped.pstex_t}}
     \caption{}
     \label{fig:ex04}
   \end{center}
 \end{figure}
 
 For the type inference of $\lam x.(x^{(2)}\ (x^{(1)}\ \lam
 w.w)):((o\to o)\to (o\to o))\to (o\to o)$, the algorithm applies the
 usual rule for abstractions seen above (\ref{eq:synt_lamx}), but in
 this case there are two instances of the bound variable $x$.  Here
 comes to work the function $\mathcal{C}$, whose rules are the
 following.

 \begin{gather}
 \label{eq:c1}
 \frac{}{\mathcal{C}(\Theta) = \emptyset}
\end{gather}
\begin{gather}
 \label{eq:c2}
 \frac{ \Unif(!^{n_1+\cdots+n_h}\Theta_1,\Theta_2,\ldots,\Theta_k)=A
   }{\mathcal{C}(!^{n_1+\cdots+n_h}\Theta_1,\ldots,\Theta_k)=\left\{\begin
       {array}{c}
       n_1+\cdots+n_h \ge 1\\
       A
     \end{array}\right.}
 \end{gather}
 Therefore the contraction of $k$ general \eal-types is obtained by
 unification and the constraint that the contracted types have at
 least one ``!''  (since in \eal{} contraction is allowed only for
 exponential formulas).
 
 Coming back to our example, the algorithm adds $c_3$ boxes passing
 through the critical point and $b_6$ boxes around the body of the
 abstraction.  The \Bang{} function modifies the first constraint in
 Equation~\eqref{eq:mod_constr2}:
 \begin{equation}
   \fbox{\ensuremath{p_{17}=b_5+p_{27}-c_3}}.
 \end{equation}
 Then the algorithm contracts the types of $x$:
 \begin{multline}
   \mathcal{C}\left(\begin{array}{c}
       b_6+b_5(b_4(p_{25}\lin p_{25})\lin p_{27}(p_{18}\lin p_{19})),\\
       b_6+c_3(p_{17}(p_{18}\lin p_{19})\lin p_{20}(p_{21}\lin p_{22}))
     \end{array}\right) = \\
   = \fbox{\ensuremath{\left\{\begin{array}{l}
           b_6+b+5\ge 1\\
           b_5 = c_3\\
           b_4=p_{17}\\
           p_{18}=p_{19} = p_{21} = p_{22} =p_{25}\\
           p_{20} = p_{27}
     \end{array}\right.}}
 \end{multline}
 Finally it removes the critical point $(p_{17}=b_5+p_{27}-c_3,\lbrack
 x^{(1)}\rbrack)$.
 
\begin{figure}
   \begin{center}
     \scalebox{.5}{
       \input{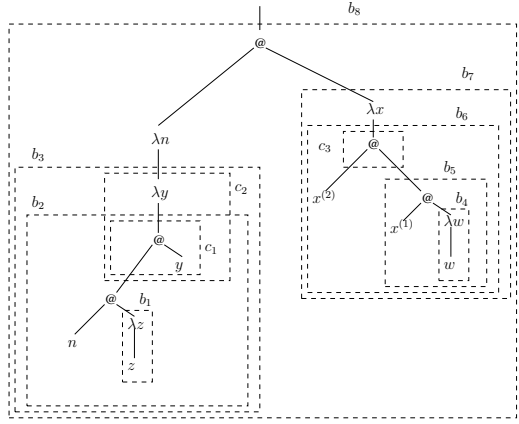}
       }
     \caption{Final superimposed decoration.}
     \label{fig:ex06}
   \end{center}
 \end{figure}

 The derivation obtained, whose decoration is shown in
 Figure~\ref{fig:ex05}, is:
 \begin{multline}
   \vdash \lam x.(x\ (x\ \lam w.w)): b_6+b_5(b_4(p_{18}\lin p_{18})\lin
   p_{20}(p_{18}\lin p_{18}))\\\lin b_6+b_5+p_{20}(p_{18}\lin p_{18}).
 \end{multline}
 \begin{figure}
   \begin{center}
     \scalebox{.7}{%
     \input{lam_x.x_x_lam_w.wealtyped.pstex_t}}
     \caption{}
     \label{fig:ex05}
   \end{center}
 \end{figure}
 The algorithm process now the whole term $(\lam n.\lam y.((n\ \lam
 z.z)\ y)\ \lam x.(x\ (x\ \lam w.w))):o\to o$.  It adds $b_7$ boxes
 around the argument of the application and unifies the \eal-types for
 the correct application:
 \begin{multline}\label{eq:last_constraint}
   \Unif\left(\begin{array}{c}     b_3+b_2(b_1(p_3(p_4\lin     p_5)\lin
       p_3(p_4\lin p_5))\lin
       p_9(p_{10}\lin p_{11}),\\
       b_7(b_6+b_5(b_4(p_{18}\lin     p_{18})\lin     p_{20}(p_{18}\lin
       p_{18}))\lin b_6+b_5+p_{20}(p_{18}\lin p_{18})
   \end{array}\right) =\\
 = \fbox{\ensuremath{\left\{\begin{array}{l}
         b_7=b_3+b_2\\
         b_1=b_6+b_5\\
         b_4=p_3=p_{20}\\
         p_4=p_5=p_{10}=p_{11}=p_{18}\\
         p_9=b_6+b_5+p_{20}
       \end{array}\right.}}
 \end{multline}
 Since this is the complete term, the final step of the algorithm is a
 single call to the function $\cors{S}$, which in this case simply
 adds $b_8$ boxes around the term. Therefore, the simply typed lambda
 term
 \begin{equation}
   (\lam n.\lam y.((n\ \lam z.z)\ y)\ \lam x.(x\ (x\ \lam w.w))):o\to o
 \end{equation}
 has \eal-type
 \begin{equation}
   !^{b_8+b_3+c_2}(!^{b_2+c_1+p_4}o\lin!^{b_2+c_1+p_4}o)
 \end{equation}
 for any $p_1,\ldots,p_{30},b_1,\ldots,b_8,c_1,c_2,c_3\in\bbbn$
 solutions of the set of constrains\footnote{We have boxed the
   constraints which were not modified by \Bang{} until the end of the
   type inference process in the exposition above. They are now all
   collected in the set of constraints~\eqref{eq:constraints}.}  in
 equations~\eqref{eq:first_constraint}--\eqref{eq:last_constraint}:
 \begin{equation}\label{eq:constraints}
   \left\{\begin{array}{l}
       b_6+b_5\ge 1\\
       b_7=b_3+b_2\\
       b_1=p_2=b_6+b_5\\
       b_5=c_3\\
       p_1=p_{16}=p_{23}=0\\
       p_9=c_1+c_2=b_6+b_5+b_4\\
       p_{17}=b_5+p_{27}-c_3\\
       b_4=p_3=p_6=p_{12}=p_{17}=p_{20}=p_{24}=p_{27}\\
       p_4=p_5=p_7=p_8=p_{10}=p_{11}=p_{13}=p_{14}=p_{15}=p_{18}\\
       p_4=p_{19}=p_{21}=p_{22}=p_{25}=p_{26}=p_{28}=p_{29}=p_{30}.
 \end{array}\right.
 \end{equation}
 The final decoration is shown in Figure~\ref{fig:ex06}.
 Considering the set of constraints in Equation~\eqref{eq:constraints}
 and the incompatibility of $c_2$ and $b_2$ stated above, the simply
 typed term 
 \begin{displaymath}
(\lam n.\lam y.((n\ \lam z.z)\ y)\ \lam x.(x\ (x\ \lam
 w.w))):o\to o
\end{displaymath}
can be typed in \eal{} either:
 \begin{enumerate}
 \item for any $n_1,\ldots,n_6\in\bbbn, n_1\ge 1$ with \eal-type
   $!^{n_3+n_5}(!^{n_1+n_2+n_4+n_6}o\lin!^{n_1+n_2+n_4+n_6}o)$ and
   decoration shown in Figure~\ref{fig:ex07a}, or
 \begin{figure}
   \begin{center}
     \scalebox{.5}{
       \input{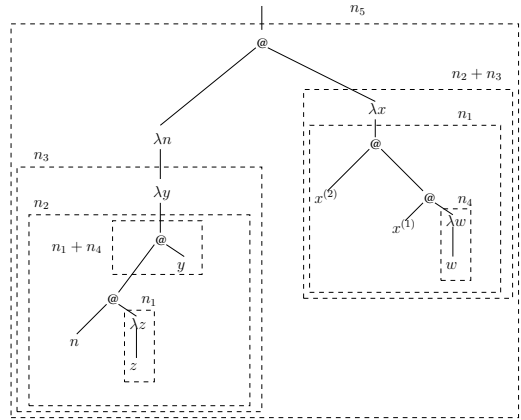}
       }
     \caption{Final decoration.}
     \label{fig:ex07a}
   \end{center}
 \end{figure}
\item for any $m_1,\ldots,m_7\in\bbbn, m_1\ge 1\ \land\ 
  m_2+m_3=m_1+m_5$ with \eal-type\\
  $!^{m_3+m_4+m_6}(!^{m_2+m_7}o\lin!^{m_2+m_7}o)$ and decoration shown
  in Figure~\ref{fig:ex07b}.
 \begin{figure}
   \begin{center}
     \scalebox{.5}{
       \input{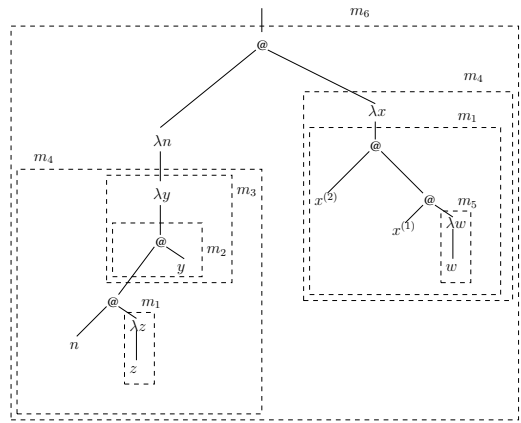}
       }
     \caption{Another possible final decoration.}
     \label{fig:ex07b}
   \end{center}
 \end{figure}
 \end{enumerate}

 \subsection{The full algorithm}
\label{section:type-inf-rules}

We define in this section the formal rules for the algorithm.
An almost complete trace of its application
to a simply typed term with no EAL type can be found in 
the Appendix.

 \begin{definition}\emph{(Type Synthesis Algorithm)}
   \label{def:typeSynthesisAlgorithm}
   Given a simply typeable lambda term $M:\sigma$, the \emph{type
     synthesis algorithm} $\cors{S}(M:\sigma)$ returns a triple
   $\langle\Theta,B,A\rangle$, where $\Theta$ is a general EAL-type,
   $B$ is a base (\emph{i.e.}, a multi-set of pairs variable, general
   EAL-type) and $A$ is a set of linear constraints.
 \end{definition}
 
 In the following $n,n_1,n_2$ are always fresh variables, $o$ is the
 base type.  Moreover, we consider $!^{n_1}(!^{n_2}\Theta)$
 syntactically equivalent to $!^{n_1+n_2}\Theta$.  
 \begin{notation}
   Given a set of linear constraints $A$ and a solution $X$ of $A$,
   for any general \eal-type $\Theta$ and for any base $B=\{x_1:
   \Theta_1, \ldots, x_n: \Theta_n \}$, we denote with $X(\Theta)$ the
   instantiation of $\Theta$ with $X$ and with $X(B)$ the
   instantiation of $B$ with $X$, \emph{i.e.}, $X(B)=\{ x_1: X(\Theta_1),
   \ldots, x_n : X(\Theta_n)\}$.
 \end{notation}

 \subsubsection{Unification: \Unif}
 \label{subsec:unification}
 
 Unification takes a set of $h\ge 2$ general \eal-types having the
 same underlying intuitionistic shape and returns a set of linear
 equations $A$ such that for any solution $X$ of $A$, the
 instantiations of the $h$ general \eal-types are syntactically
 identical.
 \begin{gather}
 \label{eq:u1}
 \frac{}{\Unif( !^{\sum n_{i_1}}o, \ldots !^{\sum n_{i_h}}o) =
   \left\{\begin{array}{l}
       \sum n_{i_1} - \sum n_{i_2} = 0\\
       \vdots\\
       \sum n_{i_{h-1}} - \sum n_{i_h} = 0
     \end{array}\right.
   }\\[2em]\displaybreak[0]
 \label{eq:u2}
 \frac{\Unif(\Theta_{1_1},\ldots,\Theta_{1_h})=A_1 \hfill
   \Unif(\Theta_{2_1},\ldots,\Theta_{2_h})=A_2
   }{\Unif\left(\begin{array}{l}
       !^{\sum n_{i_1}}(\Theta_{1_1}\linear\Theta_{2_1}),\\
       \vdots,\\
       !^{\sum n_{i_h}}(\Theta_{1_h}\linear\Theta_{2_h})
     \end{array}
   \right) = \left\{\begin{array}{c}
       \sum n_{i_1} - \sum n_{i_2} = 0\\
       \vdots\\
       \sum n_{i_{h-1}} - \sum n_{i_h} = 0\\
       A_1\\
       A_2
     \end{array}\right.}
 \end{gather}

 \subsubsection{Contraction ($\mathcal{C}$) and Type Processing (\Proc)}
 \label{subsec:contraction}
Contraction in \eal{} is allowed only for exponential formulas. Thus,
given $k$ general \eal-types, $\mathcal{C}$ returns the same set of
constraints of $\Unif$ with the additional constraint that the number
of external ! must be greater than zero.
 \begin{gather}
 \label{eq:contraction1}
 \frac{}{\mathcal{C}(\Theta) = \emptyset}\\[2em]\displaybreak[0]
 \label{eq:contraction2}
 \frac{ \Unif(!^{n_1+\cdots+n_h}\Theta_1,\Theta_2,\ldots,\Theta_k)=A
   }{\mathcal{C}(!^{n_1+\cdots+n_h}\Theta_1,\ldots,\Theta_k)=\left\{
   \begin{array}{c}   
       n_1+\cdots+n_h \ge 1\\
       A
     \end{array}\right.}
 \end{gather}
Given a simple type $\tau$, \Proc{} returns the most general \eal{}-type
whose cancellation is $\tau$, obtained by adding everywhere $p$ exponentials
(every $p$ is a fresh variable).
 \begin{gather}
 \label{eq:p1}
 \frac{}{\Proc(o)= !^p o}\\[2em]\displaybreak[0]
 \label{eq:p2}
 \frac{\Proc(\sigma)= \Theta \hfill \Proc(\tau)= \Gamma
   }{\Proc(\sigma\to\tau)= !^p (\Theta \linear \Gamma)}
 \end{gather}

 \subsubsection{Boxing: \Bang{} and \Boxing}
 \label{appendix:subsec:boxing}
The boxing procedure \Bang{} superimposes all boxes due to the presence of 
critical points. Recall the notion of slice (Definition~\ref{def:slice}) and Notation~\ref{notation:slice}.
\Bang{} has no effect if there is no critical point:
 \begin{gather}
   \frac{}{\Bang(B,\Gamma,\emptyset,A)=\langle
     B,\Gamma,A\rangle}
\end{gather}
For any slice $sl$, \Bang{} adds $c$ boxes around the subterm above
the critical points belonging to $sl$:
\begin{gather}
  \label{eq:bang}
   \frac{
   \begin{array}{l}
   \Bang\left(B_1,!^c \Gamma,cpts,A_2\right) = \langle B,\Delta,A_1\rangle\\
   B_1=\left\{x_i:\left\{\begin{array}{ll}
         !^c\Theta_i & \qquad x_i\notin sl\\
         \Theta_i & \qquad x_i\in sl
         \end{array}\right.\right\}_i
   \\
   A_2 = \left(\left\{\begin{array}{ll}
         A^j &\qquad A^j\notin sl\\
         A^j-c &\qquad A^j \in sl
         \end{array}\right.\right)_j
   \end{array}
   }{\Bang(\{x_i:\Theta_i\}_i,\Gamma,\{sl\}\cup cpts,A)=\langle
   B,\Delta,A_1\rangle}
\end{gather}
Function \Boxing{} is the wrapper for \Bang{}. It calls \Bang{} and then 
adds $b$ external boxes:
\begin{gather}
 \frac{}{\Boxing(x,B,\Gamma,cpts,A)=\langle B,\Gamma,A\rangle}\\[2em]
 \label{eq:boxing}
 \frac{ \Bang(B,\Gamma,cpts,A)=\langle B_1, \Gamma_1, A_1
   \rangle}{\Boxing(M,B,\Gamma,cpts,A)=\langle !^b B_1, !^b \Gamma_1,
   A_1 \rangle}
 \end{gather}

 \begin{proposition}\label{prop:boxing}
   Let $b,c_1,\ldots,c_k$ be the fresh variables introduced by
   $\Boxing(M,B,\Gamma,cpts,A)=\langle !^b B_1, !^b \Gamma_1, A_1
   \rangle$ and let $X$ be a solution of $A$, then
   \begin{enumerate}
   \item $X_1=(X,b=0,c_1=0,\ldots,c_k=0)$ is a solution of $A_1$;
   \item $X_1(\Gamma_1)=X(\Gamma)$;
   \item $X_1(B_1)=X(B)$.
   \end{enumerate}
 \end{proposition}
\begin{proof}
  \begin{enumerate}
  \item By Equation~\eqref{eq:bang}, for every variable $c_i$
    introduced by $\Bang$, there is a constraint $\pm
    n_{i_1}\pm\cdots\pm n_{i_{k_i}}=0$ that is changed in $\pm
    n_{i_1}\pm\cdots\pm n_{i_{k_i}}- c_i=0$, hence trivially, if the
    first one is solvable, then the second one is solvable too
    imposing $c_i=0$.  Moreover, by Equation~\eqref{eq:boxing}, $b$ is
    not added to the set of constraint, hence the thesis.
  \item By Equation~\eqref{eq:bang}
  $\Gamma_1=!^{c_1+\cdots+c_k}\Gamma$.
  \item By Equation~\eqref{eq:bang} if $B=\{x_i:\Theta_i\}_i$ then
  $B_1=\{x_i:!^{\sum_{j\in J_i} c_j}\Theta_i\}_i$ where
  $J_i\subseteq\{1,\ldots, k\}$.
\end{enumerate}
\end{proof}

 \subsubsection{Product union: $\doublecup$}
 \label{appendix:subsec:productunion}
 Product union computes all possible combinations of critical points.
 It is the culprit for the exponential complexity of the
 algorithm.
 \begin{gather}
   \frac{}{\emptyset\doublecup cpts=cpts\doublecup\emptyset =cpts} \\[2em]
   \frac{ \vertarray{sl_{2_1}}{sl_{n_1}} \doublecup
     \vertarray{sl_{1_2}}{sl_{n_2}} = cpts }{\vertarray{ sl_{1_1}}{
       sl_{n_1}} \doublecup \vertarray{ sl_{1_2}}{ sl_{n_2}} =
     \{sl_{1_1}, sl_{1_1}\cup sl_{1_2}, \ldots, sl_{1_1}\cup
     sl_{n_2}\}\cup cpts}
 \end{gather}

 \subsubsection{Type synthesis: \Sint}
 \label{subsec:typesynthesis}
 \Sint{} is the main function of the algorithm. It is defined by cases
 on the structure of the $\lam$-term. Its main cases have already been
 discussed in Section~\ref{subsec:example}. Define $\lnot app(M)$ iff
 the term $M$ is not an application.

\noindent Variable case, see equation~(\ref{eq:syntvariable}):
 \begin{gather}
   \frac{ \Proc(\sigma)=\Theta }{\Sint(x:\sigma)=\langle \Theta,
     \{x:\Theta\}, \emptyset,
     \emptyset\rangle}
\end{gather}
First abstraction case: in $\lam x.M$, $x\in\FV(M)$, see
equation~(\ref{eq:synt_lamx}):
\begin{gather}
   \frac{
     \begin{array}{l}
       h\ge 1 \;\;\;\;\; x\in\FV(M)\\
       \mathcal{C}(\Theta_1,\ldots,\Theta_h)=A_3\\
       \Boxing\left(M, B_1, \Gamma_1, cpts\cup
       \vertarray{sl_1(x)}{sl_k(x)}, A_1\right)=\left\langle B\cup
       \vertarray{x:\Theta_1}{x:\Theta_h}, \Gamma, A_2 \right\rangle\\ 
       \Sint(M:\tau)=\left\langle
       \Gamma_1,B_1, A_1,cpts\cup \orizarray{sl_1(x)}{sl_k(x)} \right\rangle\\
     \end{array}
     }{\Sint(\lam x.M:\sigma\to\tau)=\left\langle
       \Theta_1\linear\Gamma, B, \left\{\begin{array}{c}A_2\\
           A_3\end{array}\right., cpts \right\rangle}
 \end{gather}
 Second abstraction case: in $\lam x.M$, $x\notin\FV(M)$ and $M$ is an
 application:
 \begin{equation}
 \frac{
   \begin{array}{l}
     x\notin\FV((M_1\ M_2))\\
     cpts=cpts_1 \cup \{(\sum n_{i}-n=0,\mathtt{FVO}(M_1\ M_2))\}\\ 
     \Proc(\sigma)=\Theta\\
     \Boxing\left((M_1\ M_2),B_1,\Gamma_1,cpts_1,A_1
     \right)=\left\langle B, !^{\sum n_{i}}\Gamma, A \right\rangle\\
     \Sint((M_1\ M_2):\tau)=\langle \Gamma_1, B_1, A_1,
     cpts_1 \rangle 
   \end{array}
   }{\Sint(\lam    x.(M_1\    M_2):\sigma\to\tau)=\left\langle   \Theta
     \linear!^n   \Gamma,   B,   \left\{\begin{array}{c}   A\\   \sum
         n_{i}-n=0\end{array}\right., cpts \right\rangle}
 \end{equation}
 Third abstraction case: in $\lam x.M$, $x\notin\FV(M)$ and $M$ is not
 an application:
 \begin{equation}
 \frac{
   \begin{array}{l}
     \lnot app(M)\\
     x\notin\FV(M)\\
     \Proc(\sigma)=\Theta\\
     \Boxing\left(M,B_1,\Gamma_1,cpts,A_1 \right)=\left\langle
     B,\Gamma,A\right\rangle\\
     \Sint(M:\tau)=\langle \Gamma_1, B_1, A_1,
     cpts \rangle 
   \end{array}
   }{\Sint(\lam x.M:\sigma\to\tau)=\left\langle  \Theta \linear \Gamma,
     B, A, cpts \right\rangle}
 \end{equation}
 First application case: in $(M\ N)$, neither $M$ nor $N$ are
 applications, see equation~(\ref{eq:synt_app}):
 \begin{equation}
 \frac{
   \begin{array}{l}
     \lnot app(M)\ \land\  \lnot app(N)\\
     \Unif(\Theta_1,\Theta_3)=A_4\\
     \Boxing(N,B_2,\Theta_2,cpts_2,A_2)=\langle
     B_3,\Theta_3,A_3\rangle\\
     \Sint(N:\sigma)=\langle \Theta_2,B_2,A_2,cpts_2\rangle\\
     \Sint(M:\sigma\to\tau)=\langle!^{\sum n_{i}}(\Theta_1\linear
     \Gamma), B_1,A_1,cpts_1\rangle
   \end{array}}{
   \Sint((M\     N):\tau)=\left\langle     \Gamma,     B_1\cup     B_3,
     \left\{\begin{array}{c}                         A_1\\A_3\\A_4\\\sum
         n_{i}=0\end{array}\right.,                     cpts_1\doublecup
     cpts_2\right\rangle }
 \end{equation}
Second application case: in $(M\ N)$, $M$ is not an application:
 \begin{equation}
 \frac{
   \begin{array}{l}
     \lnot app(M)\\
     cpts = cpts_1\doublecup \left(cpts_2 \cup
     \{(A^1_4,\mathtt{FVO}((N_1\ N_2)))\}\right)\\ 
     \Unif(\Theta_3,\Theta_1)=A_4\\
     \Boxing((N_1\ N_2),B_2,\Theta_2,cpts_2,A_2)=\langle
     B_3,\Theta_3,A_3\rangle\\
     \Sint((N_1\ N_2):\sigma)=\langle \Theta_2,B_2,A_2,cpts_2\rangle\\
     \Sint(M:\sigma\to\tau)=\langle!^{\sum n_{i}}(\Theta_1\linear
     \Gamma), B_1,A_1,cpts_1\rangle
   \end{array}}{
   \Sint((M\   (N_1\  N_2)):\tau)=\left\langle  \Gamma,   B_1\cup  B_3,
     \left\{\begin{array}{c}                         A_1\\A_3\\A_4\\\sum
         n_{i}=0\end{array}\right., cpts\right\rangle }
 \end{equation}
 Notice that $A^1_4$ indicates the equality constraints between the
 outermost number of ! in the type of $(N_1\ N_2)$ and in the function
 part of the type of $M$.

\noindent Third application case: in $(M\ N)$, $N$ is not an
 application, see equation~(\ref{eq:synt_appappm}): 
 \begin{equation}
 \frac{
   \begin{array}{l}
     \lnot app(N)\\
     cpts = \left(cpts_1\cup \{(\sum n_{i} = 0,\mathtt{FVO}((M_1\
       M_2)))\}\right)\doublecup cpts_2\\ 
     \Unif(\Theta_1,\Theta_3)= A_4\\
     \Boxing(N,B_2,\Theta_2,cpts_2,A_2)=\langle
     B_3,\Theta_3,A_3\rangle\\
     \Sint(N:\sigma)=\langle \Theta_2, B_2, A_2, cpts_2 \rangle\\ 
     \Sint((M_1\ M_2):\sigma\to\tau)=\langle !^{\sum
       n_{i}}(\Theta_1\linear\Gamma), B_1, A_1, cpts_1\rangle  
   \end{array}
   }{\Sint((M_1\   M_2)\   N):\tau)   =   \left\langle   \Gamma,B_1\cup
     B_3,\left\{\begin{array}{c}A_1\\A_3\\A_4\\\sum                n_{i}
         =0\end{array}\right., cpts\right \rangle}
 \end{equation}
Fourth application case: in $(M\ N)$, both $M$ and $N$ are applications:
 \begin{equation}
 \frac{
   \begin{array}{l}
     cpts_4 = cpts_2\cup \{(A_4^1,\mathtt{FVO}((N_1\ N_2)))\}\\
     cpts_3 = cpts_1\cup \{(\sum n_{i} = 0,\mathtt{FVO}((M_1\ M_2)))\}\\
     \Unif(\Theta_3,\Theta_1)= A_4\\
     \Boxing((N_1\ N_2),B_2,\Theta_2,cpts_2,A_2)=\langle
     B_3,\Theta_3,A_3\rangle\\
     \Sint((N_1\ N_2):\sigma)=\langle \Theta_2, B_2, A_2, cpts_2 \rangle\\
     \Sint((M_1\ M_2):\sigma\to\tau)=\langle !^{\sum
     n_{i}}(\Theta_1\linear\Gamma), B_1, A_1, cpts_1\rangle  
   \end{array}
   }{\Sint((M_1\ M_2)\ (N_1\  N_2)):\tau) = \left\langle \Gamma,B_1\cup
     B_3,\left\{\begin{array}{c}A_1\\A_3\\A_4\\\sum
         n_{i}=0\end{array}\right.,       cpts_3\doublecup       cpts_4
   \right\rangle}
 \end{equation}

 \subsubsection{Type synthesis algorithm: $\cors{S}$}
\label{sec:type-synth-algor}
  $\cors{S}$ is the top level call for the algorithm. It passes
the call to \Sint{}, takes its result, boxes the term, forgets the 
critical points and eventually contracts the common variables
in the base. 
 \begin{equation}\label{eq:synt}
   \frac{
   \begin{array}{l}
     \mathcal{C}(\Theta_{1_1},\ldots,\Theta_{k_1})=A_1
     \qquad \ldots \qquad
     \mathcal{C}(\Theta_{1_h},\ldots,\Theta_{k_h})=A_h\\
     \Boxing(M,B_1,\Theta_1,cpts,A')=\left\langle
       \vertarray{x_1:\Theta_{1_1},\ldots,
     x_1:\Theta_{k_1},}{x_h:\Theta_{1_h}, \ldots, x_h:\Theta_{k_h}} 
        ,\Theta,A\right\rangle\\
     \Sint(M:\sigma)=\langle\Theta_1, B_1, A', cpts\rangle
   \end{array}}{
   \cors{S}(M:\sigma)=\left\langle   \Theta,   \left\{x_1:\Theta_{1_1},
       x_2:\Theta_{1_2},        \ldots,       x_h:\Theta_{1_h}\right\},
     \left\{\begin{array}{c}
         A\\A_1\\\vdots\\A_h\end{array}\right.\right\rangle }
 \end{equation}

\section{Properties of the type inference algorithm}
\label{sect:properties}
We will prove in this section that our algorithm $\cors{S}$ is complete with
respect to the notion of $\Gamma \vdashEAL M: A$ introduced in
Definition~\ref{def:officialEALtyping}.  Correctness and completeness
of $\cors{S}$ are much simpler if, instead of EAL, we formulate proofs
and results with reference to an equivalent natural deduction
formulation, discussed in the following subsection. Before, we state
the obvious fact that our algorithm does not loop, since any rule
$\Sint$ decreases the structural size of the $\lam$-term $M$, any rule
$\Unif$ decreases the size of the type
$\Theta$ and any rule $\Bang$ and $\doublecup$ decreases the size of
the set of critical points $cpts$.

 \begin{proposition}[Termination]
   Let $M$ be a simply typed term and let $\sigma$ its simple
   type. $\cors{S}(M:\sigma)$ always
   terminates with a triple $\langle \Theta, B, A\rangle$.
 \end{proposition}
 
 The algorithm is exponential in the size of the $\lambda$-term,
 because to investigate all possible derivations we need to (try to)
 box all possible combinations of critical points (see the clauses for
 the product union, $\doublecup$,
 in Section~\ref{appendix:subsec:productunion}), that are roughly
 bounded by the size of the term.
 
 \subsection{NEAL}
 \label{sec:NEAL}
 The natural deduction calculus (NEAL) for EAL in given in
 Figure~\ref{fig:NEAL},
 after~\cite{Asperti:1998:lcs,Benton:Bierman:dePaiva:Hyland:1993:tlca:lncs,Roversi:1998:asian:lncs}.

 \begin{lemma}[Weakening]\label{lem:neal-weak}
   If $\Gamma\vdashNEAL A$ then $B,\Gamma\vdashNEAL A$.
 \end{lemma}

 \begin{figure}
   \begin{center}
     \begin{tabular}{| c c |}
       \hline &\\
       $\infer[ax]{\Gamma,A\vdashNEAL A}{}$ &
       $\infer[contr]{\Gamma,\Delta\vdashNEAL B}{
       \Gamma\vdashNEAL !A & \Delta,!A,!A\vdashNEAL B}$\mbox{\hspace{1em}}\\&\\
       \mbox{\hspace{1em}} 
       $\infer[(\linear I)]{\Gamma\vdashNEAL A\linear
       B}{\Gamma,A\vdashNEAL B}$ & $\infer[(\linear
       E)]{\Gamma,\Delta\vdashNEAL B}{\Gamma\vdashNEAL A\linear B &
       \Delta\vdashNEAL A}$\\&\\
       \multicolumn{2}{| c |}{
       $\infer[!]{\Gamma,\Delta_1,\ldots,\Delta_n\vdashNEAL
       !B}{\Delta_1\vdashNEAL !A_1
       \cdots \Delta_n\vdashNEAL !A_n & A_1,\ldots,A_n\vdashNEAL
       B}$}\\
       &\\
       \hline
     \end{tabular}
     \caption{Natural Elementary Affine Logic in sequent style
       notation}
     \label{fig:NEAL}
   \end{center}
 \end{figure}
 
 To annotate NEAL derivations, we use terms generated by the following
 grammar (\emph{elementary affine terms} $\Lambda^{EA}$):
 $$M::=x\ |\ \lam x.M\ |\ (M\ M)\ |\ \promote{M}{M}{x}{M}{x}\ |\ 
 \contr{M}{M}{x}{x}$$
 
 Observe that in $\promote{M}{M}{x}{M}{x}$, the $[{}^{M}/x]$ is a kind of
 explicit substitution.  To define ordinary substitution, define first
 the set of free variables of a term $M$, $\mathtt{FV}(M)$,
 inductively as follows:
 \begin{itemize}
 \item $\mathtt{FV}(x)=\{x\}$
 \item $\mathtt{FV}(\lam x.M)=\mathtt{FV}(M)\smallsetminus\{x\}$
 \item $\mathtt{FV}(M_1\ M_2)=\mathtt{FV}(M_1)\cup\mathtt{FV}(M_2)$
 \item $\mathtt{FV}(\promote{M}{M_1}{x_1}{M_n}{x_n})=
   \bigcup_{i=1}^{n}\mathtt{FV}(M_i)\cup\mathtt{FV}(M)\smallsetminus
   \{x_1,\ldots,x_n\}$
 \item $\mathtt{FV}(\contr{M}{N}{x_1}{x_2})=
   (\mathtt{FV}(M)\smallsetminus\{x_1,x_2\})\cup\mathtt{FV}(N)$
 \end{itemize}
 
 Ordinary substitution $N\{M/x\}$ of a term $M$ for the free
 occurrences of $x$ in $N$, is defined in the obvious way:
 \begin{enumerate}
 \item $x\{M/x\} = M$;
 \item $y\{M/x\} = y$ if $y\neq x$;
 \item $\lam x.N\{M/x\} = \lam x.N$;
 \item $\lam y.N\{M/x\} = \lam z.(N\{z/y\}\{M/x\})$ where $z$ is a
   fresh variable;
 \item\label{point:substapp} $(N\ P)\{M/x\} = (N\{M/x\}\ P\{M/x\})$;
 \item\label{point:substpromotealphaconvert}
   $\promote{N}{P_1}{x_1}{P_n}{x_n}\{M/x\} =\\
   \promote{N\{y_1/x_1\}\cdots\{y_n/x_n\}\{M/x\}}{P_1\{M/x
     \}}{y_1}{P_n\{M/x\}}{y_n}$\\ if $x\notin\{x_1,\ldots,x_n\}$,
   where $y_1,\ldots,y_n$ are all fresh variables;
 \item $\promote{N}{P_1}{x_1}{P_n}{x_n}\{M/x\} =
   \promote{N}{P_1\{M/x\}}{x_1}{P_n\{M/x\}}{x_n}$\\ if $\exists i$
   s.t.  $x_i = x$;
 \item\label{point:substcontralphaconvert} $\contr{N}{P}{y}{z}\{M/x\}
   = \contr{N\{y'/y\}\{z'/z\}\{M/x\}}{P\{M/x\}}{y'}{z'}$ if
   $x\notin\{y,z\}$, where $y',z'$ are fresh variables;
 \item $\contr{N}{P}{y}{z}\{M/x\} = \contr{N}{P\{M/x\}}{y}{z}$ if
   $x\in\{y,z\}$.
 \end{enumerate}
 
 Elementary terms may be mapped to $\lam$-terms, by forgetting the
 exponential structure:
 \begin{itemize}
 \item $x^* =x$
 \item $(\lam x.M)^* = \lam x.M^*$
 \item $(M_1\ M_2)^* = (M_1^*\ M_2^*)$
 \item $(\promote{M}{M_1}{x_1}{M_n}{x_n})^* =
   M^*\{M_1^*/x_1,\ldots,M_n^*/x_n\}$
 \item $(\contr{M}{N}{x_1}{x_2})^* = M^*\{N^*/x_1,N^*/x_2\}$
 \end{itemize}

 \begin{definition}\emph{(Legal elementary terms)}
   The elementary terms are \emph{legal} under the following
   conditions:
 \begin{enumerate}
 \item $x$ is legal;
 \item $\lam x.M$ is legal iff $M$ is legal;
 \item $(M_1\ M_2)$ is legal iff $M_1$ and $M_2$ are both legal and
   $\mathtt{FV}(M_1)\cap\mathtt{FV}(M_2)=\emptyset$;
 \item $\promote{M}{M_1}{x_1}{M_n}{x_n}$ is legal iff $M$ and $M_i$
   are legal for any $i\quad 1\le i\le n$ and
   $\mathtt{FV}(M)=\{x_1,\ldots,x_n\}$ and $\left(i\neq j\Rightarrow
     \mathtt{FV}(M_i)\cap\mathtt{FV}(M_j) =\emptyset\right)$;
 \item $\contr{M}{N}{x}{y}$ is legal iff $M$ and $N$ are both legal
   and $\mathtt{FV}(M)\cap\mathtt{FV}(N)=\emptyset$.
 \end{enumerate}
 \end{definition}

 \begin{proposition}
   If $M$ is a legal term, then every free variable
   $x\in\mathtt{FV}(M)$ is linear in $M$.
 \end{proposition}
 \begin{proof}
   By trivial induction on the structure of $M$ using definitions of
   legal terms and $\mathtt{FV}$.
 \end{proof}

 \begin{note}
   From now on we will consider only legal terms.
 \end{note}

 \begin{notation}
   Let $\Gamma=\{x_1:A_1,\ldots,x_n:A_n\}$ be a basis.
   $dom(\Gamma)=\{x_1,\ldots,x_n\}$; $\Gamma(x_i)=A_i$;
   $\Gamma\upharpoonright V=\{x:A|x\in V \land A=\Gamma(x)\}$.
 \end{notation}
 
 The term assignment system is shown in
 Figure~\ref{fig:termassignment}, where all bases in the premises of
 the contraction, $\linear$ elimination and !-rule, have domains with
 empty intersection.
 \begin{figure}
   \begin{center}
     \scalebox{.8}{%
     \fbox{%
     \begin{tabular}{ c c }
       &\\
       $\infer[ax]{\Gamma,x:A\vdashNEAL x:A}{}$ &
       $\infer[contr]{\Gamma,\Delta\vdashNEAL \contr{N}{M}{x}{y}:B}{
       \Gamma\vdashNEAL M:!A & \Delta,x:!A,y:!A\vdashNEAL N:B}$
       \mbox{\hspace{1em}}\\&\\ 
       \mbox{\hspace{1em}} $\infer[(\linear I)]{\Gamma\vdashNEAL \lam
       x.M:A\linear 
       B}{\Gamma,x:A\vdashNEAL M:B}$ & $\infer[(\linear
       E)]{\Gamma,\Delta\vdashNEAL (M\ N):B}{\Gamma\vdashNEAL M:A\linear B &
       \Delta\vdashNEAL N:A}$\\&\\
       \multicolumn{2}{ c }{
       $\infer[!]{\Gamma,\Delta_1,\ldots,\Delta_n\vdashNEAL
       \promote{N}{M_1}{x_1}{M_n}{x_n}:!B}{\Delta_1\vdashNEAL M_1:!A_1
       \cdots \Delta_n\vdashNEAL M_n:!A_n & x_1:A_1,\ldots,x_n:A_n\vdashNEAL
       N:B}$}\\
       &\\
     \end{tabular}}}
     \caption{Term Assignment System for Natural Elementary Affine
       Logic}
     \label{fig:termassignment}
   \end{center}
 \end{figure}

 \begin{lemma}\mbox{}
 \begin{enumerate}
 \item If $\Gamma\vdashNEAL M:A$ then $\mathtt{FV}(M)\subseteq
   dom(\Gamma)$;
 \item if $\Gamma\vdashNEAL M:A$ then $\Gamma\upharpoonright
   \mathtt{FV}(M)\vdashNEAL M:A$.
 \end{enumerate}
 \end{lemma}

 \begin{lemma}[Substitution]
   If $\Gamma,x:A\vdashNEAL M:B$ and $\Delta\vdashNEAL N:A$ and
   $dom(\Gamma)\cap dom(\Delta)=\emptyset$ then
   $\Gamma,\Delta\vdashNEAL M\{N/x\}:B$.
 \end{lemma}
 \begin{proof}
   Recalling that both $M$ and $N$ are legal terms, by easy induction
   on the structure of $M$.
 \end{proof}

 \begin{theorem}[Equivalence]
   $\Gamma\vdash_{EAL} A$ if and only if $\Gamma\vdash_{NEAL} A$.
 \end{theorem}
 \begin{proof}
   \begin{itemize}
   \item[(if)] By induction, using the cut rule. It is also
   possible to prove, by an easy inspection of the cut-elimination
   theorem for \eal{}, that it is possible to eliminate just the
   exponential cuts, leaving the logical ones.
   \item[(only if)] The only interesting case is $(\lin L)$. The
   proof is identical to the case of intuitionistic logic.
   \end{itemize}
 \end{proof}

 \begin{lemma}[Unique Derivation]
   For any legal term $M$ and formula $A$, if there is a valid
   derivation of the form $\Gamma\vdashNEAL M:A$, then such derivation
   is unique (up to weakening).
 \end{lemma}
 
 A notion of reduction is needed to state and obtain completeness of
 the type inference algorithm.  We define two
 \emph{logical} reductions ($\to_\beta$ and $\to_{\mathsf{dup}}$)
 corresponding to the elimination of principal cuts in EAL. The other
 five reductions are permutation rules, allowing contraction to be
 moved out of a term.
 \begin{displaymath}
 \begin{array}{l}
 (\lam x.M\ N) \qquad \to_\beta \qquad M\{N/x\}\\\\
 \contr{N}{\promote{M}{M_1}{x_1}{M_n}{x_n}}{x}{y} \qquad
 \to_{\mathsf{dup}} \\ 
 \qquad \contr{\contr{N\{ ^{\promote{M}{x_1'}{x_1}{x_n'}{x_n}}/x\}
     \{ ^{\promote{M'}{y_1'}{y_1}{y_n'}{y_n}}/y\}}{M_1}{x_1'}{y_1'}
   \cdots}{M_n}{x_n'}{y_n'}\\\\
 !(M)[ ^{M_1}/x_1,\cdots,
 ^{\promote{N}{P_1}{y_1}{P_m}{y_m}}/x_i,\cdots, ^{M_n}/x_n] \qquad
 \to_{!-!}\\ 
 \qquad  !(M\{N/x_i\})[ ^{M_1}/x_1,\cdots, ^{P_1}/y_1, \cdots, ^{P_m}/y_m,
 \cdots ^{M_n}/x_n]\\\\
 (\contr{M}{M_1}{x_1}{x_2}\ N) \qquad \to_{@-\mathsf{c}}
 \qquad \contr{(M\{x_1'/x_1,x_2'/x_2\}\ N)}{M_1}{x_1'}{x_2'}\\\\
 (M\ \contr{N}{N_1}{x_1}{x_2}) \qquad \to_{@-\mathsf{c}}
 \qquad \contr{(M\ N\{x_1'/x_1,x_2'/x_2\})}{N_1}{x_1'}{x_2'}\\\\
 !(M)[ ^{M_1}/x_1,\cdots,
 ^{\contr{M_i}{N}{y}{z}}/x_i,\cdots, ^{M_n}/x_n] \qquad
 \to_{!-c}\\
 \qquad \contr{!(M)[ ^{M_1}/x_1,\cdots, ^{M_i\{y'/y,z'/z\}}/x_i,\cdots,
   ^{M_n}/x_n]}{N}{y'}{z'}
 \end{array}
 \end{displaymath}
 \begin{displaymath}
 \begin{array}{l}
 \contr{M}{\contr{N}{P}{y_1}{y_2}}{x_1}{x_2} \qquad
 \to_{\mathsf{c}-\mathsf{c}} \qquad
 \contr{\contr{M}{N\{y_1'/y_1,y_2'/y_2\}}{x_1}{x_2}}{P}{y_1'}{y_2'}\\\\
 \lam x.\contr{M}{N}{y}{z} \qquad \to_{\lam-\mathsf{c}} \qquad
 \contr{\lam x.M}{N}{y}{z} \mbox{\textrm{ where }}
 x\notin\mathtt{FV}(N)
 \end{array}
 \end{displaymath}
 where $M'$ in the $\to_{\mathsf{dup}}$-rule is obtained from $M$
 replacing all its free variables with fresh ones ($x_i$ is replaced
 with $y_i$); $x_1'$ and $x_2'$ in the $\to_{@-\mathsf{c}}$-rule, $y'$
 and $z'$ in the $\to_{!-c}$-rule and $y_1',y_2'$ in the
 $\to_{\mathsf{c}-\mathsf{c}}$-rule are fresh variables.

 \begin{definition}
   The reduction relation on legal terms $\rightsquigarrow$ is defined
   as the reflexive and transitive closure of the union of $\to_\beta,
   \to_{\mathsf{dup}}, \to_{!-!},
   \to_{@-\mathsf{c}},\to_{!-\mathsf{c}}, \to_{\mathsf{c}-\mathsf{c}},
   \to_{\lam-\mathsf{c}}$.
 \end{definition}

 \begin{proposition}
   Let $M\rightsquigarrow N$ and $M$ be a legal term, then $N$ is a
   legal term.
 \end{proposition}

 \begin{proposition}~\label{prop:*}
   Let $M\!\!\to_r\! N$ where $r$ is not $\to_\beta$, then $M^* =
   N^*$.
 \end{proposition}

 \begin{lemma}~\label{lemma:normalforms}
   Let $M$ be a well typed term in
   $\{\mathsf{dup},!-!,@-\mathsf{c},!-\mathsf{c},\mathsf{c}-\mathsf{c},\lam-
   \mathsf{c}\}$-normal form, then
   \begin{enumerate}
   \item if $R=\contr{N}{P}{x}{y}$ is a subterm of $M$, then either
     $P=(P_1\ P_2)$ or $P$ is a variable;
   \item if $R=\promote{N}{P_1}{x_1}{P_k}{x_k}$ is a subterm of $M$,
     then for any $i\in\{1,\ldots,k\}$ either $P_i=(Q_i\ S_i)$ or
     $P_i$ is a variable.
   \end{enumerate}
 \end{lemma}

 \begin{theorem}[Subject Reduction]\label{th:subj-red}
   Let $\Gamma\vdashNEAL M:A$ and $M\rightsquigarrow N$, then
   $\Gamma\vdashNEAL N:A$.
 \end{theorem}

 \subsection{Properties of the Type Inference Algorithm}

 The following Lemma states that any slice in the set of critical
 points bars the rest of the term.
 \begin{lemma}\label{lem:critical-points-paths}
   Let $\Sint(M:\sigma)=\langle \Theta,B,A,cpts\rangle$. For any slice
   $sl$ in $cpts$, $sl=\{cpt_1,\ldots,cpt_k\}$, for every path from
   the root of the syntax tree of $M$ to any leaf, there exists at
   most one $cpt_i$ in the path. 
 \end{lemma}
 \begin{proof}
   By induction on $M$. The unique interesting case is $M=(M_1\ M_2)$.
   The thesis holds by inductive hypothesis and by a simple inspection
   of rules for $\Sint$ and for the product union.
 \end{proof}
The following lemma illustrates the relation between the set of critical
points calculated by the algorithm for a given term $M$ and a
particular class of decompositions of $M$.
\begin{lemma}\label{lem:critical-points-decomposition}
   Let $\Sint(M:\sigma)=\langle \Theta,B,A,cpts\rangle$.
   \begin{enumerate}
   \item $\forall \{cpt_1,\ldots,cpt_k\}=sl\in cpts$ there exist
     $P,(N_{1_1}\ N_{2_1}),\ldots,(N_{1_k}\ N_{2_k})$ such that $P$ is
     not a variable, $x_1,\ldots,x_k\in\FV(P)$ and $M=P\{(N_{1_1}\
     N_{2_1})/x_1,\ldots,(N_{1_k}\ N_{2_k})/x_k\}$;
   \item $\forall P,(N_{1_1}\ N_{2_1}),\ldots,(N_{1_k}\ N_{2_k})$ such
     that $P$ is not a variable, $x_1, \ldots, x_k \in \FV(P)$ and $M
     = P \{(N_{1_1}\ N_{2_1})/x_1,\ldots,(N_{1_k}\ N_{2_k})/x_k\}$,
     there exists $\{cpt_1,\ldots,cpt_k\}=sl\in cpts$ such that
     $cpt_i$ is the critical point at the root of $(N_{1_i}\ 
     N_{2_i})$.
   \end{enumerate}
 \end{lemma}
 \begin{proof}
   By structural induction on $M$.
   \begin{enumerate}
   \item If $M$ is a variable, the thesis trivially holds being
     $cpts=\emptyset$. If $M=\lam x.M'$, either $sl$ consists of a
     single critical point corresponding to the root of $M'$, then
     $P=\lam x.y$, or $sl$ is a slice of $M'$, then by inductive
     hypothesis there exists $P'$ s.t. the thesis holds for $M'$. We
     take $P=\lam x.P'$. Finally if $M=(M_1\ M_2)$, if in $sl$ there
     is a critical point $cpt_i$ corresponding to the root of $M_1$
     then by Lemma~\ref{lem:critical-points-paths} all the other
     critical points in $sl$ belong to $M_2$ or there is only one
     critical point corresponding to the root of $M_2$. In the first
     case by inductive hypothesis there exists $P_2$ s.t. the thesis
     holds for $M_2$ and $sl$ without $cpt_i$. Then we take $P=(y\
     P_2)$. The other cases are analogous.
   \item If $M$ is a variable then $\not\exists P$ and the thesis
     trivially holds. If $M=\lam x.M'$ then $P=\lam x.P'$. If $P'$ is
     a variable, then the slice to consider is the one containing only
     the critical point corresponding to the root of $M'$. Such a
     slice has been added to $cpts$ in the rule for $\Sint(\lam
     x.(M_1\ M_2):\sigma)$ where $x\notin\FV((M_1\ M_2))$. Otherwise
     the thesis holds by inductive hypothesis. Finally if $M=(M_1\
     M_2)$, then $P=(P_1\ P_2)$. If both $P_1$ and $P_2$ are not a
     variable, then by inductive hypothesis there exists $sl_1$ and
     $sl_2$. Then the thesis holds by definition of product union. The
     other cases are analogous.
   \end{enumerate}
 \end{proof}

Consider the \emph{length} $L(M)$ of an \eal-term $M$ defined
inductively:
    \begin{align*}
      L(x) &= 0\\
      L(\lam x.M) &= 1 + L(M)\\
      L((M\ N)) &= 1 + L(M) + L(N)\\
      L(\promote{M}{M_1}{x_1}{M_n}{x_n}) &= L(M)  + \sum_{i=1}^{n} L(M_i)\\
      L(\contr{M}{N}{x}{y}) &= L(M) + L(N).
    \end{align*}
\begin{definition}
  An \eal-term $M$ is \emph{simple} if and only if
    \begin{enumerate}
    \item $M$ has no subterm of the form $\contr{M_1}{M_2}{x}{y}$
      where $(M_2)^*$ is not a variable,
    \item $L(M)=L((M)^*)$
    \end{enumerate}
  \end{definition}
  \begin{fact}
    A simple \eal-term contracts at most variables.
  \end{fact}
  \begin{definition}\label{def:candidate}
    The set of \emph{candidate \eal-terms} is the set of all
    \eal-terms $P$ such that
    \begin{enumerate}
    \item\label{def:candidate-1} $P$ is in
      $\{!-!,@-\mathsf{c},!-\mathsf{c},\mathsf{c}-\mathsf{c},
      \lam-\mathsf{c},\mathsf{dup}\}$-normal form;
    \item\label{def:candidate-2} $P$ is simple;
    \item\label{def:candidate-3} if $\contr{R}{Q}{x}{y}$ is a subterm
      of $P$, then $x,y\in \FV(R)$;
    \item\label{def:candidate-4} if $\promote{R}{Q_1}{x_1}{Q_k}{x_k}$
      is a subterm of $P$, then $R$ is not a variable.
    \end{enumerate}
  \end{definition}
  \begin{definition}
    Given a general \eal-type $\Theta$ we define its \emph{erasure}
    $\overline\Theta$ as the simple type obtained by $\Theta$ erasing
    all the exponentials ``!'' and changing $\lin$ into $\to$.
  \end{definition}

  \begin{lemma}\label{lem:proc-eal-type}
    For any $\Theta$ general \eal-type there exists $X$
    s.t. $X(\Proc(\overline{\Theta}))=\Theta$. 
  \end{lemma}
 \begin{theorem}[Completeness]
 \label{theor:completeness}
 Let $\Gamma\vdashNEAL P:\Psi$ and let $P$ be a candidate \eal-term.
 Let $\cors{S}(P^*:\overline\Psi)=\langle\Theta,B,A\rangle$, then
 there exists $X$ integer solution of $A$ such that $X(B)\subseteq
 \Gamma$, $\Psi =X(\Theta)$ and $X(B)\vdashNEAL P:X(\Theta)$.
 \end{theorem}
 \begin{proof}
    By induction on $P$.
   \begin{itemize}
   \item If $\Gamma,x:\Psi\vdashNEAL x:\Psi$ then
     $\mathcal{S}(x:\overline{\Psi})=\langle \Proc(\overline{\Psi}),\{x:
     \Proc(\overline{\Psi})\}, \emptyset\rangle$ and the
     thesis holds by Lemma~\ref{lem:proc-eal-type} being any $X$
     solution of the empty set of constraints.
   \item If the type derivation ends with
     \begin{displaymath}
       \infer{\Gamma,\Delta\vdashNEAL \contr{N}{x}{y}{z}:\Psi}{%
       \Gamma\vdashNEAL x:!\Phi
       &
       \Delta,y:!\Phi,z:!\Phi\vdashNEAL N:\Psi}
     \end{displaymath}
     then the thesis
       holds by inductive hypothesis on
       $\Delta,y:!\Phi,z:!\Phi\vdashNEAL N:\Psi$. 
   \item If $P$ is an abstraction then the type derivation is
     \begin{displaymath}
       \infer{\Gamma\vdashNEAL \lam x.M: \Psi \linear \Phi}{%
         \Gamma,x:\Psi\vdashNEAL M:\Phi}
     \end{displaymath}
     The thesis holds by inductive hypothesis. Notice that the
     solution $X$ instantiates all variables introduced by the
     \Boxing{} call of the rule for $\mathcal{S}$ to 0. It is easy to
     see looking at the rules for \Boxing{} that if in the solution
     $X$ there is one variable introduced by \Boxing{} that is not set
     to zero, then the type is exponential and $\Psi \linear \Phi$ is
     not.
   \item If $P$ is an application
     \begin{displaymath}
       \infer{\Gamma,\Delta\vdashNEAL (M\ N):\Psi}{%
         \Gamma\vdashNEAL M:\Phi\linear\Psi
         &
         \Delta\vdashNEAL N:\Phi
         }
     \end{displaymath}
     By inductive hypothesis there are solutions $X_1$ for $M$ and
     $X_2$ for $N$. Now, by the same considerations of the previous
     point, $X_1$ sets all variables introduced by the last \Boxing{}
     call to 0. Thus the constraint $\sum n_j=0$ of the rule for
     \Sint{} is satisfied. Moreover $X_1,X_2$ satisfies
     the constraints for the unification of types, because they are
     identical by hypothesis. Hence the thesis holds.
   \item Finally, if the derivation is
     \begin{displaymath}
       \infer{\Gamma,\Delta_1,\ldots,\Delta_n\vdashNEAL
         \promote{N}{M_1}{x_1}{M_n}{x_n}:!\Psi}{%
         \Delta_1\vdashNEAL M_1:!\Phi_1\cdots\Delta_n\vdashNEAL
         M_n:!\Phi_n
         &
         x_1:\Phi_1,\ldots,x_n:\Phi_n\vdashNEAL N:\Psi}
     \end{displaymath}
     then by Lemma~\ref{lemma:normalforms} either $M_i$ is a variable
     or an application. If all $M_i$ are variables, then the thesis
     holds getting the solution of the inductive hypothesis and
     increasing the variable $b$ introduced by \Boxing{} by one.
     
     If there is an $M_i$ that is an application, then by
     Lemma~\ref{lem:critical-points-decomposition} there is a critical
     point collected by the algorithm at the root of $M_i$. Then
     we take as solution $X$ the union of the solutions obtained by
     inductive hypothesis with the variable introduced by \Boxing{}
     for the critical point corresponding to $M_i$ increased by one.
   \end{itemize}
\end{proof}

In the statement of the previous theorem, the request on the $\{!-!,
@-\mathsf{c}, !-\mathsf{c}, \mathsf{c}-\mathsf{c}, \lam-\mathsf{c},
\mathsf{dup}\}$-normal form is not a loss of generality, for the
subject reduction lemma and Proposition~\ref{prop:*}.  By
Lemma~\ref{lemma:normalforms}, the only restriction induced by the
request of contracting at most variable is the exclusion of elementary
terms with subterms of the form $\contr{R}{(Q_1\ Q_2)}{x}{y}$ or
$!(R)[{}^{P_1}/ x_1, \cdots, {}^{(Q_1\ Q_2)}/ x,\cdots,{}^{P_n}/ x_n]$
with $\contr{S}{x}{y}{z}$ subterm of $R$.  Recalling the discussion at
the end of Section~\ref{sec:eal}, we see that these terms, in a sense,
``contract too much'' --- in the sharing graph of the corresponding
$\lam$-term $P^*$, there would be fan nodes corresponding to
non-variable contractions.  We also do not take into account
elementary affine terms with ``false contractions''. This is not a
limitation by Lemma~\ref{lem:neal-weak} and Theorem~\ref{th:subj-red}.
Finally we discard term such $!(x)[M/x]$. Again this is not a
limitation, in fact $(!(x)[M/x])^*=M^*$ and $\Gamma\vdashNEAL
!(x)[M/x]:!\Psi$ if and only if $\Gamma\vdashNEAL M:!\Psi$.

\begin{notation}
  We use
   \begin{displaymath}
     \infer{\Gamma\vdash !^n(N)[{}^{M}/x]:!^n B}{%
       \Gamma\vdash M:!^n A
       &
       x:A\vdash N:B}
   \end{displaymath}
   as a shorthand for
   \begin{displaymath}
     \infer{\Gamma\vdash \overbrace{!(\cdots!(}^{n}N)[{}^{x_1} /
     x]\cdots)[{}^{M} / x_{n-1}]:\overbrace{!\cdots!}^{n}B}{%
       \Gamma\vdash M:\overbrace{!\cdots!}^{n}A
       &
       \infer*{x_{n-1}:\overbrace{!\cdots!}^{n-1}: A \vdash
       \overbrace{!(\cdots!(}^{n-1}N)[{}^{x_1} / x]\cdots)[{}^{x_{n-1}} /
       x_{n-2}]:\overbrace{!\cdots!}^{n-1}B}{%
         \infer{}{%
           x_2:!!A\vdash x_2:!!A
           &
           \infer{x_1:!A\vdash !(N)[{}^{x_1}/x]:!B}{%
             x_1:!A\vdash x_1:!A
             &
             x:A\vdash N:B}
           }
         }
       }
   \end{displaymath}
\end{notation}
\begin{lemma}[Superimposing of derivations]
 \label{lemma:superposition}
 Let $\cors{S}(M:\sigma)=\langle \Theta,B,A\rangle$ and let $A$ be
 solvable. If there is a solution $X_1$ of $A$ that instantiates two
 boxes belonging to two superimposed derivations that are not
 compatible, then there exists another solution $X_2$ where all the
 instantiated boxes belong to the same derivation.
 
 Moreover $X_1(\Theta)=X_2(\Theta)$ and $X_1(B)=X_2(B)$.
 \end{lemma} 
 \begin{proof}
   The proof of the lemma can be easily understood if we follow the
   intuition explained below with an example.

   We may think of boxes as levels; boxing a subterm can then
   be seen as raising that subterm, as in
   Figure~\ref{fig:3Dboxedterm}, where also some types label the edges
   of the syntax tree of a simple term.  In particular, the edge
   starting from the $@$-node and ending in $x_0$ has label
   $!^{n_2}(\alpha\linear!^{n_1}(\beta\linear \gamma))$ at level $0$
   (nearest to $x_0$) and has label
   $(\alpha\linear!^{n_1}(\beta\linear \gamma))$ at level $n_2$. This
   is the graphical counterpart of the !-rule
   $$\infer[!^{n_2}]{\ldots,x_0:!^{n_2}T,\ldots\vdash
     \ldots}{\ldots,x_0:T,\ldots\vdash \ldots}$$
 \begin{figure}
   \begin{center}
     \scalebox{.9}{%
     \input{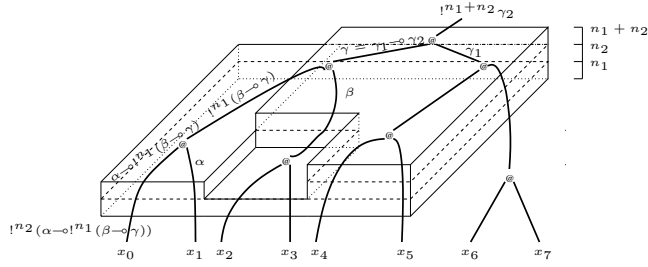}}
     \caption{Boxes as levels.}
     \label{fig:3Dboxedterm}
   \end{center}
 \end{figure}
 The complete decoration of Figure~\ref{fig:3Dboxedterm} can be
 produced in NEAL in two ways: by the instantiation of
 \begin{displaymath}
 !^{n_2}\left((((x_0\  x_1)y)((x_4\ x_5)w))\right)[(x_2\ x_3)/y,(x_6\ 
 x_7)/w]
 \end{displaymath}
 and\footnote{The correct legal terms should have all free variable
   inside the square brackets. We omit to write variables when they
   are just renamed, for readability reasons (compare the first
   elementary term above with the (fussy) correct one
   $!^{n_2}\left((((x_0\ x_1)y)((x_4\ x_5) w) ) \right)[x_0'/x_0,
   x_1'/x_1, (x_2\ x_3)/y, x_4'/x_4, x_5'/x_5, (x_6\ x_7)/w]$).}
 \begin{displaymath}
 !^{n_1}\left(((z (x_2\  x_3)) ((x_4\ x_5)w)) \right)[(x_0\ x_1)/z, (x_6\ 
 x_7)/w],
 \end{displaymath}
 which are boxes belonging to two different derivations.  Graphically
 such an instantiation can be represented as in the first row of
 Figure~\ref{fig:3Dboxes}, where incompatibility is evident by the
 fact that the boxes are not well stacked, in particular the
 rectangular one covers a hole.  To have a correct EAL-derivation it
 is necessary to find the equivalent, well stacked configuration (that
 corresponds to the subsequent application of boxes from the topmost
 to the bottommost).

 \begin{figure}
 \begin{center}
   \scalebox{.9}{%
   \input{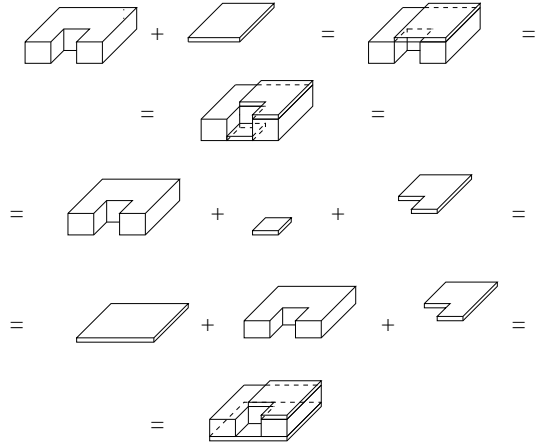}}
 \caption{Equivalences of boxes.}
 \label{fig:3Dboxes}
 \end{center}
 \end{figure}
 The procedure by which we find the well stacked box configuration is
 visualized in Figure~\ref{fig:3Dboxes}.  The reader may imagine the
 boxes subject to gravity (the passage from the first to the second
 row of Figure~\ref{fig:3Dboxes}) and able to fuse each other when
 they are at the same level (the little square in the third row fuse
 with the solid at its left in the passage from the third to the
 fourth row).
 
 The ``gravity operator'' corresponds to finding the minimal common
 subterm of all the superimposed derivations and it is useful for
 finding the correct order of application of the ! rule.  The ``fusion
 operator'' corresponds to the elimination of a cut between two
 exponential formulas.  Moreover, the final configuration of
 Figure~\ref{fig:3Dboxes} corresponds to a particular solution of the
 set of constraints produced by the type synthesis algorithm, that
 instantiates the following boxes:
 \begin{displaymath}
 !^{n_1}\left(!^{n_2-n_1}\left(!^{n_1}\left(((z\ y) ( (x_4\ x_5) w)
       )\right)[(x_0\   x_1)/z] \right) [(x_2\  x_3)/y] \right) (x_6\
 x_7)/w] 
 \end{displaymath}
 
 Finally, notice that during the procedure all types labeling the
 boundary edges of the lambda-term never changes, \emph{i.e.},  the
 instantiations of the term type (the label of the topmost edge) and
 the base types (the labels of the edges at the bottom) remain
 unchanged.
 
 Now let $\mathcal{S}(M:\sigma)=\langle \Theta,B,A\rangle$ and let $X$
 be the solution that instantiates $k$ overlapping---thus
 incompatible---boxes. Consider the boxed syntax tree of $M$ and
 associate to any node its level, \emph{i.e.}, the number of boxes
 containing the node. Notice that if there is a wire connecting tho
 nodes $a$ of level $\ell$ and $b$ of level $\ell+k$, then the type
 labeling the wire is $!^k\Psi$ near $a$ and $\Psi$ near $b$,
 \emph{i.e.}, the sum of level and number of exponentials for types
 labeling the syntax tree is an invariant. We break the boxes using
 the following procedure: starting from the root of the syntax tree of
 $M$, we are at level $i=0$; we proceed with a breath first visit and
 whenever encounter a node of level $\ell\neq i$ we close $i$ boxes,
 open $\ell$ boxes and set $i$ to $\ell$.
 
 \begin{figure}
   \centering
   \scalebox{.9}{%
   \includegraphics[width=.55\textwidth]{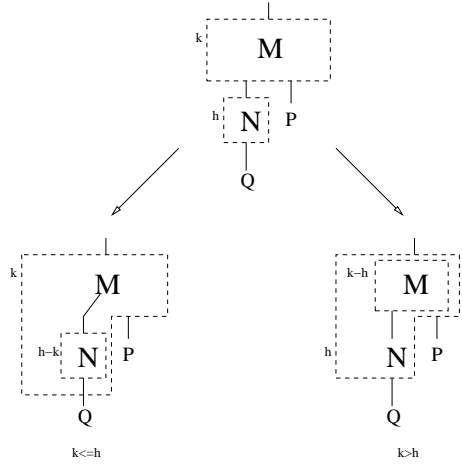}}
   \caption{Fusion of boxes.}
   \label{fig:box-fusion}
 \end{figure}
 At the end of the procedure described above there are no more
 overlapping boxes, but it could be happen that there is a variable
 $x$ not in the same boxes of its binding lambda node. Such
 configuration of boxes is not correct. However the level of the
 variable and lambda node is the same because the procedure of
 breaking boxes does not change level of nodes. Moreover all nodes
 belonging to the path from the lambda node to the variable have level
 higher or equal to the level of the variable since they all were
 initially in the same box and some of them were eventually also in
 some overlapping boxes that increase the level. Hence we can fuse
 boxes until variable and corresponding binder are in the same box.
 The fusion operation is shown in Figure~\ref{fig:box-fusion} and
 described by the following equation:
 \begin{displaymath}
   \begin{array}{rcl}
     && !^k(M\{{}^{!^{h-k}(N)[{}^{Q}/z]}/x\})[{}^{P}/y]\text{ if $k\le
     h$}\\
     &\nearrow &\\
     !^{k}(M)[{}^{P}/y,{}^{!^{h}(N)[{}^{Q}/z]}/x]&&\\
     &\searrow&\\
     && !^h(!^{k-h}(M)[{}^{N}/x])[{}^{Q}/z,{}^{P}/y]\text{ if $k>h$}
   \end{array}
 \end{displaymath}
 After all fusions are performed, all variables are in the same boxes
 of their lambda binders and there are no more overlapping boxes, thus
 the decoration obtained corresponds to an \eal-derivation. By
 completeness exists $X_2$ solution corresponding to such
 decoration. Moreover types labeling the syntax tree are unchanged by
 the transformations applied, hence the thesis.
 \end{proof}

 \begin{theorem}[Soundness]
 \label{theor:soundness}
 Let $\cors{S}(M:\sigma)=\langle\Theta,B,A\rangle$.  For every $X$
 integer solution of $A$, there exists $P$ candidate
 \eal-term such that $P^* = M$ and $X(B)\vdashNEAL P:X(\Theta)$.
 \end{theorem}
 \begin{proof}
   By induction on the structure of $M$, using the superimposing
   lemma.  
   \begin{figure}[!h]
     \centering
     \scalebox{.7}{%
       \input{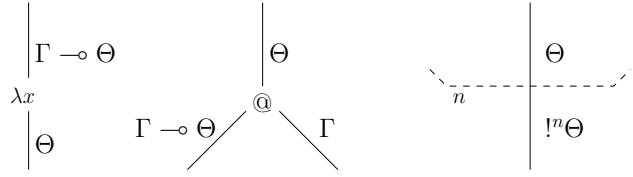}%
     }
     \caption{Type labels for decorated syntax trees.}
     \label{fig:typed-graphs}
   \end{figure}
   We first need a definition:
   \begin{definition}
     A syntax tree $T$ is \emph{correctly decorated} if the edges of
     the graph are labeled according to Figure~\ref{fig:typed-graphs}
     (in the rightmost picture, $\Theta$ is inside $n$ boxes).
     Moreover all edges connecting a variable $x$ occurring multiple,
     are labeled with the same type $!^n \Gamma$. In the case the
     variable is abstracted, the type label of variable is
     syntactically identical to the argument part of the type label of
     the edge at the root of the abstraction.
   \end{definition}
   Given a correctly decorated syntax tree, and an instantiation $X$
   for the general \eal-types labeling its edges such that the number
   of exponentials for types of multiple variables is greater than 1,
   it is easy to build the corresponding NEAL derivation, using the
   Curry-Howard isomorphism and eventually applying a contraction
   before the $\linear$ introduction for binded variables and at the
   end of the derivation for free variables.

   Thus, in order to prove soundness of our algorithm, it is
   sufficient to prove by structural induction on $M$ that we can
   build a correctly decorated syntax tree. If the solution taken into
   account instantiates two overlapping boxes we use
   Lemma~\ref{lemma:superposition}. Hence without loss of generality
   we can consider $X$ such that all boxes are compatible.  
   \begin{figure}[!h]
     \centering
     \scalebox{.7}{%
       \input{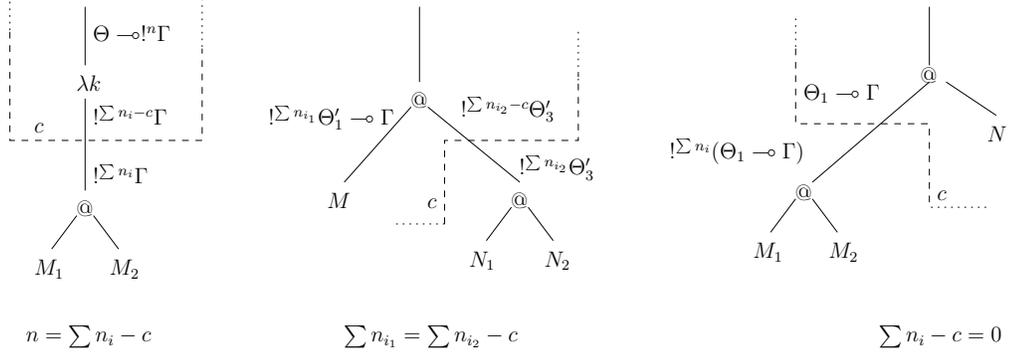}%
     }
     \caption{Decorations given by \Boxing.}
     \label{fig:boxing-critical-points}
   \end{figure}
   The only interesting part of the proof is the checking of rules for
   \Boxing. In Figure~\ref{fig:boxing-critical-points} it is shown how
   build a correctly decorated syntax tree when the solution $X$
   instantiates a box passing through a critical point (all three
   cases of critical points are depicted).
   
   Finally we need to prove that $P$ is a candidate \eal-term. Points
   2 and 3 of Definition~\ref{def:candidate} hold by construction of
   the NEAL derivation from the correctly decorated syntax tree, which
   also guarantees that $P$ is in
   $\{@-\mathsf{c},!-\mathsf{c},\mathsf{c}-\mathsf{c},
   \lam-\mathsf{c},\mathsf{dup}\}$-normal form.  Point 4 holds by
   definition of $\Boxing$, and $P$ is in !-!-normal form by the
   superimposing lemma.
 \end{proof}

\begin{theorem}[Main theorem]
  \label{theor:main}
  Let $M$ be a simply typeable $\lam$-term. For any basis $\Gamma$ and
  \eal{} formula $C$: \\
$ \Gamma \vdashEAL M:C$
  iff $\cors{S}(M:\overline C)=\langle\Theta,B,A\rangle$ and $A$
  admits an integral solution $X$ such that $X(B)\subseteq \Gamma $
  and $C = X(\Theta)$.
\end{theorem}
\begin{proof}
  ($\Rightarrow$) $\Gamma \vdashEAL M:C$ is established by a sharing
  graph where no fan node faces the root of a subgraph.
  It is ready to see that the corresponding EAL-term is a
  \emph{candidate} EAL-term. Theorem~\ref{theor:completeness} allows
  to conclude.
  
  ($\Leftarrow$) By Theorem~\ref{theor:soundness}, there is an
  EAL-term $P$ such that $P^*=M$ and $X(B) \vdashNEAL P:X(\Theta)$.
  The NEAL-term $P$ codes a sharing graph establishing $X(B) \vdashEAL
  P^*:X(\Theta)$.
\end{proof}

\begin{lemma}\label{theor:lemmatipoprinc}
  Let $M$ be a simply typeable $\lam$-term; let $\sigma$ be its
  principal type schema, and let $\tau$ be any other type for $M$. If
  $\cors{S}(M:\tau)=\langle\Theta,B,A\rangle$ and $A$ admits a
  solution $X$, then $\cors{S}(M:\sigma)=\langle\Theta',B',A'\rangle$
  and there exists $X'$ solution of $A'$.
\end{lemma}
\begin{proof}
  We have to show that it is not the case that $A$ admits a solution
  and $A'$ is unsolvable. Constraints are added only by
  contraction~\eqref{eq:contraction2} or
  unification~\eqref{eq:u2}.  The former constraints depend only on
  the structure of the syntax tree of the \emph{term} and hence they
  are not affected by the type change.  As for the latter, changing
  $\tau$ into $\sigma$ makes some unification constraints disappear.
  In fact, it is possible to decompose $\Theta$ in
  $\Theta'\{x_1\to\Sigma_1,\ldots,x_n\to\Sigma_n\}$. When the
  algorithm synthesizes $M:\sigma$, all unification constraints in $A$
  regarding $\Sigma_1\ldots\Sigma_n$ disappear, and we obtain $A'$ (up
  to renaming). In order to prove that $A'$ is $A$ minus the set of
  unification constraints produced by $\Sigma_1\ldots\Sigma_n$, it is
  sufficient to inspect the definitions of $\Proc$ and $\Unif$.  As
  the solution space has increased, it is not possible that $A'$ has
  no solution.
\end{proof}

\begin{corollary}
  Let $M$ be a simply typeable $\lam$-term and let $\sigma$ be its principal
  type schema. For any basis $\Gamma$ and \eal{} formula $C$:
  $\Gamma\vdashEAL M:C$ iff
  $\cors{S}(M:\sigma)=\langle\Theta,B,A\rangle$, $A$ admits an
  integral solution $X$ and there exists a substitution $S$ from type
  variables to \eal-types such that $S(X(B))\subseteq\Gamma$ and
  $S(X(\Theta))=C$.
\end{corollary}

The corollary gives a weak notion of principal type for EAL. Any EAL
type of a term arises as an instance of a solution of the constraints
obtained for its simple principal type schema.  The result, however,
does not say anything on the structure of these !-decorated instances.
The study of a general notion of principal schema for EAL is the
subject of~\cite{Coppola:Ronchi:2003:tlca:lncs}.  On the other hand, the
corollary is enough to establish the decidability of type inference.

\begin{theorem}
It is decidable whether, given a type-free $\lam$-term $M$,
there exist an EAL formula $C$ and a basis $\Gamma$ such that
$\Gamma \vdashEAL M:C$.
\end{theorem}

\section{Conclusions}
\label{sect:conclusions}
We have presented an algorithm for assigning EAL types to type-free,
pure $\lam$-terms, obtained as the (technically non trivial)
elaboration of the idea of ``box decoration'' of a simple type
derivation.  The algorithm is shown complete with respect to the
notion of EAL types introduced in
Definition~\ref{def:officialEALtyping}.  If we change the constraints
collected by the algorithm, the same technique can be used to obtain
linear logic derivations. Or, we may use the algorithm to infer types
for a more liberal notion of EAL-typeability.

 \subsection{Linear decorations of intuitionistic derivations}  
\label{sect:LL}
The problem to obtain linear logic derivations from intuitionistic
derivations has been thoroughly
studied~\cite{Danos:Joinet:Schellinx:1995:aml,Schellinx:1994:phd,Roversi:1992:ictcs}.
Our linear constraints method can be used to obtain a variety of such
decorations.

The implicational fragment of linear logic can be obtained from \eal{} 
by adding the rules:
   \begin{displaymath}
     \begin{array}{cc}
       \infer[\epsilon]{\Gamma, !A\vdash B}{\Gamma, A\vdash B} &
       \infer[\delta]{\Gamma, !!A\vdash B}{\Gamma, !A\vdash B}
     \end{array}
   \end{displaymath}
Introduce now the rule $(d+b)$
   \begin{displaymath}
     \infer[(d+b)]{\Gamma,!^{x-(d+b)}A\vdash B}{\Gamma,!^x A\vdash B}
     \qquad \left\{\begin{array}{l}
         x\ge 0\\
         d+b\le x-1\\
         d\ge 0\\
         -1\le b\le 0,
       \end{array}\right.
   \end{displaymath}
   which acts as a multiple $\delta$ rule, except when $d=0$ and
   $b=-1$. In this case it is the same of an $\epsilon$ rule.  It is
   easy to prove that $\Gamma\vdash_{\mathrm{LL}} B$ iff
   $\Gamma\vdash_{\mathrm{LL}-\{\delta,\epsilon\}\cup(d+b)} B$.
   Extend now the maximal decoration method as follows.  After each
   logical rule, interleave $n$ !-rules, and then, for each formula
   $A$ in the context , add one $(d_i+b_i)$ rule and $e_i$
   $\epsilon$-rules. For example
   \begin{displaymath}
     \infer{A\vdash B\linear C}{A,B\vdash C}
   \end{displaymath}
   becomes
   \begin{displaymath}
     \infer{!^{n-(d_1+b_1)+e_1}A\vdash !^{n-(d_2+b_2)+e_2}B\linear
     !^{n}C}{
     \infer=[\epsilon]{!^{n-(d_1+b_1)+e_1}A, !^{n-(d_2+b_2)+e_2}B\vdash
     !^{n}C}{
     \infer[(b+d)]{!^{n-(d_1+b_1)+e_1}A, !^{n-(d_2+b_2)}B\vdash
     !^{n}C}{
     \infer=[\epsilon]{!^{n-(d_1+b_1)+e_1}A, !^{n}B\vdash
     !^{n}C}{
     \infer[(b+d)]{!^{n-(d_1+b_1)}A, !^{n}B\vdash
     !^{n}C}{
     \infer=[!]{!^{n}A, !^{n}B\vdash
     !^{n}C}{A,B\vdash C}}}}}}
   \end{displaymath}
   During the type inference, the set of constraints obtained from
   unification and contraction is augmented by the constraints of
   rules $(d_i+b_i)$.  It is not difficult to see that any solution of
   the set of constraints collected by the algorithm gives a linear
   logic derivation having the original intuitionistic derivation as a
   skeleton.
   
   Notice that the meta-derivations obtained by the above procedure
   represents a set of LL derivations complete for the provability of
   LL formulas. In fact, the unique derivations of LL that are not
   direct instances of the previous meta-derivations are those where
   exponential rules are applied in a different order. However, it is
   easy to see that the rules under discussion may be freely permuted.
   For example, if $\Gamma\vdash_{LL}B$ with an application of !-rule
   followed by an $\epsilon$-rule, then $\Gamma\vdash_{LL}B$ with
   inverted order of exponential rules (the proof is similar for the
   other cases).
   
   The use of linear constraints allows now the use of linear
   programming techniques to obtain decorations with specific
   properties.  By minimizing the objective function $\sum_i
   n_i+\sum_j (d_j+b_j)+\sum_k e_k$, we obtain decorations using a
   minimal number of boxes. Or, we may minimize only the use of
   $\epsilon$ and $\delta$ rules, if we minimize $\sum_j
   (d_j+b_j)+\sum_k e_k$. In the language of optimal reduction, these
   are decorations introducing a minimal number of brackets and
   croissants, and are thus the natural candidates to be used as
   initial translations for those $\lam$-terms which does not have an
   EAL type.

\subsection{Arbitrary contractions}
Instead of using Definition~\ref{def:officialEALtyping}, we may would
like an algorithm complete with respect to the notion given directly
by Figure~\ref{fig:EAL}, that is, allowing arbitrary contractions (and
not only variable contractions) in the sharing graphs.  Proceed as
follows.  Given a generic sharing graph, first decompose
it into several subgraphs with the property that no fan faces a
subgraph; than readback them, obtaining a set of
lambda-terms. For example, the graph of
Figure~\ref{fig:only-typeable-with-contraction} of
Section~\ref{sec:eal} can be decomposed in $\lam z.\lam x.\lam w.(k\ 
k\ w)$ and $(x\ z)$.  After the decomposition, call the type synthesis
algorithm separately on every subterm, calculate the suitable
unification constraints with $\Unif$, collect all the constraints
in a single system, and solve it.

  \begin{figure}
   \centering
   \scalebox{.65}{%
   \includegraphics[width=.55\textwidth]{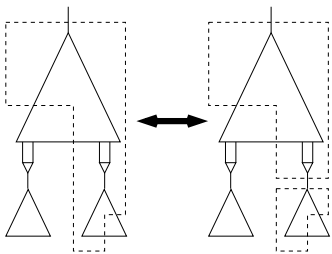}}
   \caption{Box fusion for arbitrary contractions.}
   \label{fig:box-fusion-arbitrary-contractions}
 \end{figure}
 
 This procedure computes all possible decorations, except those boxes
 that surround more than one subterm. However, the proof of the
 superimposing lemma allows to conclude that there is a
 decoration with a box around more than one subterm if and only
 if there exists a decoration with boxes only around a
 single subterm, with the same type (see
 Figure~\ref{fig:box-fusion-arbitrary-contractions} for a graphical
 intuition).

\section*{Acknowledgments}
We are happy to thank Harry Mairson, for extended comments and
criticism on previous versions of the paper; and Simona Ronchi della
Rocca, for the many discussions, suggestions, and comments.

\bibliographystyle{alpha}
\bibliography{EA-typing}
\appendix
\section{Appendix}
\label{ex:non-typeable}
We have already observed that  the simply typed lambda term
   \begin{displaymath}
     (\lam n.(n\ \lam y.(n\ \lam z.y))\ \lam x.(x\ (x\ y))):o
   \end{displaymath}
   is not typeable in \eal. 
\begin{figure}[!b]
  \centering
  \scalebox{.5}{
    \input{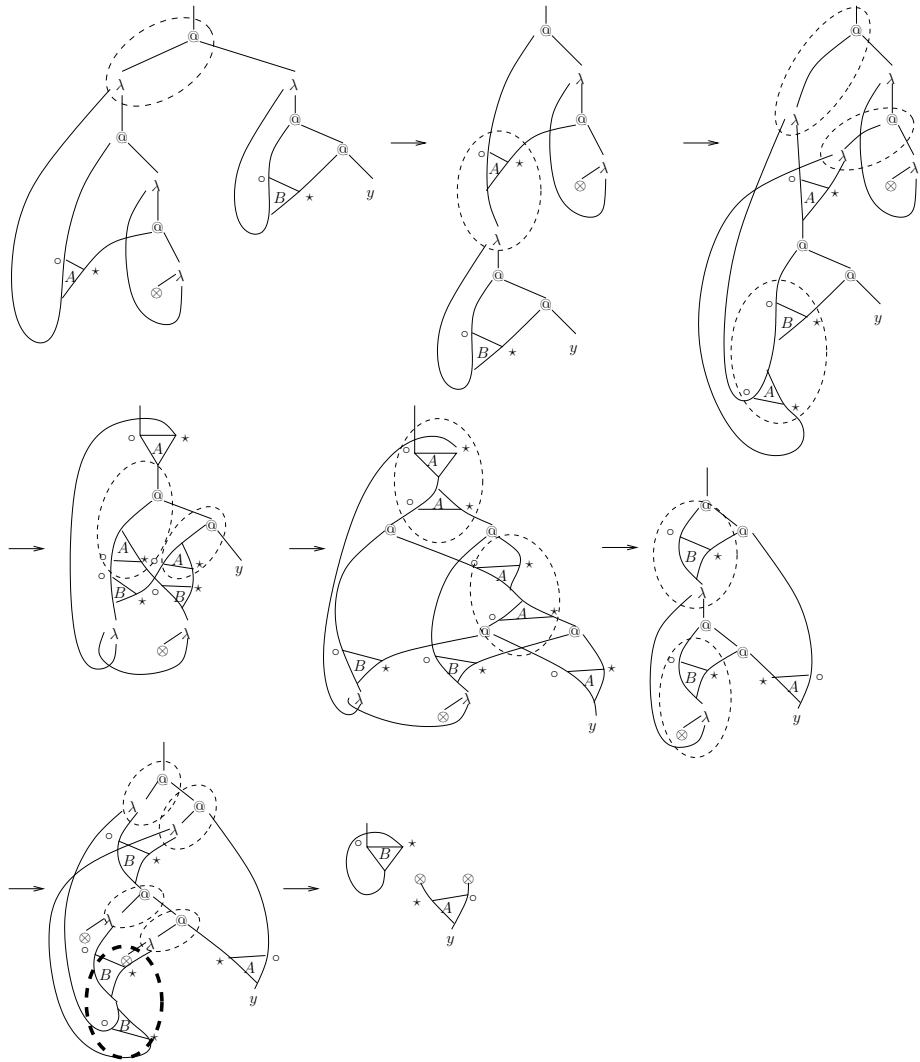}
    }
  \caption{Incorrect reduction of $(\lam  n.(n\ \lam y.(n\  \lam z.y))\  \lam
    x.(x\  (x\ y)))$.}
  \label{fig:incorrect-reduction}
\end{figure}
If one knows optimal reduction~\cite{Asperti:Guerrini:1998} , this can
be seen in a simple way, writing the term as a sharing graph and
reducing it in the abstract algorithm by matching fans by labels (see
Figure~\ref{fig:incorrect-reduction} where the redexes fired at every
step are indicated by a dashed oval).  The sharing graph in normal
form is {\em a cycle}, that is a sharing graph which does not
correspond to any $\lambda$-term (least to say to $y$, which is the
normal form of the given term).  This means that the oracle {\em is\/}
needed for the reduction of this term, and hence it cannot have a type
in \eal{}.

We can give a formal proof, by calling the type inference 
algorithm  on such a term. The following is a trace of
the execution, where each box delimits the call
and return of a single type inference rule:

\begin{xtabular}{r@{}r@{}r@{}r@{}r@{}r@{}r@{}l|l|l|l|l|}
  \Sint & \multicolumn{7}{@{}l}{$(\lam n.(n\ \lam y.(n\ \lam z.y))\
  \lam x.(x\ (x\ y))):o)$}\\
  &\Sint&\multicolumn{6}{@{}l}{$(\lam n.(n\ \lam y.(n\ \lam
    z.y)):((o\to o)\to o)\to o)$} \\
  \cline{3-12}
  &\vline & \Sint& \multicolumn{5}{@{}l}{$((n\ \lam y.(n\ \lam
  z.y)):o)$} & \multicolumn{4}{c|}{}\\
  \cline{4-8}
  &\vline &\vline & \Sint& \multicolumn{4}{@{}l|}{$(n:(o\to o)\to o)$} & \multicolumn{4}{c|}{}\\
  &\vline &\vline & & \Proc& \multicolumn{3}{@{}l|}{$((o\to o)\to
    o)=p_1(p_2(p_3\linear p_4)\linear p_5)$} & \multicolumn{4}{c|}{}\\ 
  &\vline &\vline & & \multicolumn{4}{@{}l|}{$=\langle p_1(p_2(p_3\linear
    p_4)\linear p_5),\{n:p_1(p_2(p_3\linear p_4)\linear
    p_5)\},\emptyset,\emptyset\rangle$} & \multicolumn{4}{c|}{}\\ 
  \cline{4-8}
  &\vline & & \multicolumn{5}{c}{} & \multicolumn{4}{c|}{}\\
  \cline{4-11}
  &\vline &\vline & \Sint& \multicolumn{4}{@{}l}{$(\lam y.(n \lam
  z.y):o\to o)$} & \multicolumn{3}{c|}{} &\\
  \cline{5-10}
  &\vline &\vline &\vline & \Sint& \multicolumn{3}{@{}l}{$((n\ \lam
  z.y):o)$} & \multicolumn{2}{c|}{} & &\\
  \cline{6-8}
  &\vline &\vline &\vline &\vline & \Sint& \multicolumn{2}{@{}l|}{$(n:(o\to o)\to o)$} & \multicolumn{2}{c|}{} & &\\
  &\vline &\vline &\vline &\vline & & \Proc & $((o\to o)\to o)= p_6(p_7(p_8\linear p_9)\linear
    p_{10})$ & \multicolumn{2}{c|}{} & &\\ 
  &\vline &\vline &\vline &\vline & & \multicolumn{2}{@{}l|}{$=\langle p_6(p_7(p_8\linear
    p_9)\linear p_{10}),\{n:p_6(p_7(p_8\linear p_9)\linear
    p_{10})\},\emptyset,\emptyset\rangle$} & \multicolumn{2}{c|}{} & &\\ 
  \cline{6-8}
  &\vline &\vline &\vline & & \multicolumn{3}{c}{} & \multicolumn{2}{c|}{} & &\\
  \cline{6-9}
  &\vline &\vline &\vline &\vline & \Sint& \multicolumn{2}{@{}l}{$(\lam z.y:o\to o)$} & & & &\\
  \cline{7-8}
  &\vline &\vline &\vline &\vline &\vline &\Sint& $(y:o)$ & & & &\\
  &\vline &\vline &\vline &\vline &\vline & & \Proc$(o)=p_{11}$ & & & &\\
  &\vline &\vline &\vline &\vline &\vline & & $=\langle
    p_{11},\{y:p_{11}\},\emptyset,\emptyset\rangle$ & & & &\\ 
  \cline{7-8}
  &\vline &\vline &\vline &\vline & & \multicolumn{2}{@{}l}{\Boxing$(y,\{y:p_{11}\}, p_{11},
    \emptyset,\emptyset )= \langle \{y:p_{11}\}, p_{11},
    \emptyset\rangle$} & & & &\\   
  &\vline &\vline &\vline &\vline & & \multicolumn{2}{@{}l}{\Proc$(\alpha)=p_{12}$} & & & &\\
  &\vline &\vline &\vline &\vline & & \multicolumn{2}{@{}l}{$=\langle p_{12}\linear
    p_{11},\{y:p_{11}\},\emptyset,\emptyset\rangle$} & & & &\\
  \cline{6-9}
  &\vline &\vline &\vline & & \Boxing & \multicolumn{4}{@{}l|}{$(\lam
    z.y,\{y:p_{11}\},p_{12}\linear p_{11}, \emptyset,\emptyset)$} & &\\ 
  &\vline &\vline &\vline & & & \multicolumn{4}{@{}l|}{\Bang$(\{y:p_{11}\},p_{12}\linear
    p_{11}, \emptyset,\emptyset)=\langle \{y:p_{11}\},p_{12}\linear
    p_{11}, \emptyset\rangle$} & &\\ 
  &\vline &\vline &\vline & & & \multicolumn{4}{@{}l|}{$=\langle
    \{y:b_1+p_{11}\},b_1(p_{12}\linear p_{11}), \emptyset\rangle$} & &\\ 
  &\vline &\vline &\vline & & \multicolumn{5}{@{}l|}{\Unif$(p_7(p_8\linear
    p_9),b_1(p_{12}\linear p_{11}))=\left\{\begin{array}{l}
      p_7 = b_1\\
      p_8 = p_{12}\\
      p_9 = p_{11}
    \end{array}\right.$} & &\\
  &\vline &\vline &\vline & & \multicolumn{5}{@{}l|}{$=\left\langle
      p_{10},\{n:p_6(p_7(p_8\linear p_9)\linear p_{10}),
      y:b_1+p_{11}\},\left\{\begin{array}{l}  
          p_7 = b_1\\
          p_8 = p_{12}\\
          p_9 = p_{11}\\
          p_6 = 0
        \end{array}\right.,\emptyset\right\rangle$} & &\\
  &\vline &\vline &\vline & & \multicolumn{5}{@{}l|}{$=\langle p_{10},\{n:b_1(p_8\linear
    p_9)\linear p_{10},y:b_1+p_9\},\emptyset,\emptyset\rangle$} & &\\ 
  \cline{5-10}
  &\vline &\vline & & \Boxing & \multicolumn{6}{@{}l|}{$((n\ \lam z.y),
    \{n:b_1(p_8\linear p_9)\linear p_{10},y:b_1+p_9\}, p_{10},
    \emptyset, \emptyset)$} &\\ 
  &\vline &\vline & & & \multicolumn{6}{@{}l|}{$=\langle \{n:b_2(b_1(p_8\linear
    p_9)\linear p_{10}),y:b_2+b_1+p_9\},b_2+p_{10},\emptyset\rangle$} &\\
  &\vline &\vline & & \multicolumn{7}{@{}l|}{$\mathcal{C}(b_2+b_1+p_9)=\emptyset$} &\\
  &\vline &\vline & & \multicolumn{7}{@{}l|}{$=\langle b_2+b_1+p_9\linear
    b_2+p_{10},\{n:b_2(b_1(p_8\linear p_9)\linear p_{10})\},
    \emptyset,\emptyset\rangle$} &\\ 
  \cline{4-11}
  &\vline & & & \Boxing & \multicolumn{7}{@{}l|}{$(\lam y.(n\ \lam z.y),
    \{n:b_2(b_1(p_8\linear p_9)\linear p_{10})\}, b_2+b_1+p_9\linear
    b_2+p_{10}, \emptyset, \emptyset)$}\\ 
  &\vline & & & & \multicolumn{7}{@{}l|}{$=\langle \{n:b_3+b_2(b_1(p_8\linear
    p_9)\linear p_{10})\}, b_3(b_2+b_1+p_9\linear b_2+p_{10}),
    \emptyset \rangle$}\\ 
  &\vline & & & \Unif & \multicolumn{7}{@{}l|}{$(p_2(p_3\linear
    p_4),b_3(b_2+b_1+p_9\linear b_2+p_{10}))= \left\{\begin{array}{l} 
          p_2 = b_3\\
          p_3 = b_2+b_1+p_9\\
          p_4 = b_2+p_{10}
        \end{array}\right.$}\\
  &\vline & & & \multicolumn{8}{@{}l|}{$=\left\langle p_5, \left\{\begin{array}{l}
      n:p_1(p_2(p_3\linear p_4)\linear p_5),\\
      n:b_3+b_2(b_1(p_8\linear p_9)\linear p_{10})
    \end{array}\right\},\left\{\begin{array}{l}
          p_2 = b_3\\
          p_3 = b_2+b_1+p_9\\
          p_4 = b_2+p_{10}\\
          p_1 = 0
        \end{array}\right.,\emptyset\right\rangle$}\\
  &\vline & & & \multicolumn{8}{@{}l|}{$=\left\langle
      p_5,\left\{\begin{array}{l}  
          n:b_3(b_2+b_1+p_9\linear b_2+p_{10})\linear p_5,\\
          n:b_3+b_2(b_1(p_8\linear p_9)\linear p_{10})
        \end{array}\right\},\emptyset,\emptyset\right\rangle$}\\
  \cline{3-12}
  & & & \Boxing & \multicolumn{4}{@{}l}{$\left((n\ \lam y.(n\ \lam z.y)),\left\{\begin{array}{l}
      n:b_3(b_2+b_1+p_9\linear b_2+p_{10})\linear p_5,\\
      n:b_3+b_2(b_1(p_8\linear p_9)\linear p_{10})
    \end{array}\right\},p_5,\emptyset,\emptyset\right)$}\\
  & & & & \multicolumn{4}{@{}l}{$=\left\langle \left\{\begin{array}{l}
      n:b_4(b_3(b_2+b_1+p_9\linear b_2+p_{10})\linear p_5),\\
      n:b_4+b_3+b_2(b_1(p_8\linear p_9)\linear p_{10})
    \end{array}\right\},b_4+p_5,\emptyset\right\rangle$}\\
  & & & $\mathcal{C}$ & \multicolumn{4}{@{}l}{$(b_4(b_3(b_2+b_1+p_9\linear b_2+p_{10})\linear
  p_5), b_4+b_3+b_2(b_1(p_8\linear p_9)\linear p_{10}))$}\\
  & & & & \multicolumn{4}{@{}l}{$=\left\{\begin{array}{l}
      b_4\ge 1\\
      b_4 = b_4+b_3+b_2\\
      b_3 = b_1\\
      b_2+b_1+p_9 = p_8\\
      b_2+p_{10} = p_9\\
      p_5 = p_{10}
    \end{array}\right.\qquad=\left\{\begin{array}{l}
      b_4\ge 1\\
      b_3 = 0\\
      b_2 = 0\\
      b_1 = 0\\
      p_8 = p_5\\
      p_9 = p_5\\
      p_{10} = p_5
    \end{array}\right.$}\\
  & & & \multicolumn{5}{@{}l}{$=\langle b_4((p_5\linear p_5)\linear
    p_5)\linear b_4+p_5, \emptyset,\{b_4\ge 1\},\emptyset\rangle$}\\
  \cline{2-10}
  &\vline & \Sint & \multicolumn{5}{@{}l}{$(\lam x.(x\ (x\ y)):(o\to o)\to
    o)$}&\multicolumn{2}{c|}{}\\
  \cline{4-9}
  &\vline &\vline & \Sint & \multicolumn{4}{@{}l}{$((x\ (x\ y)):o)$}&&\\
  &\vline &\vline & & \Sint & \multicolumn{3}{@{}l}{$(x:o\to o)$}&&\\
  &\vline &\vline & & & \multicolumn{3}{@{}l}{$=\langle p_1(p_2\linear p_3),\{x:p_1(p_2\linear p_3)\},
  \emptyset,\emptyset\rangle $}&&\\ 
  &\vline &\vline & & \Sint & \multicolumn{3}{@{}l}{$((x\ y):o)$}&&\\
  &\vline &\vline & & & \Sint & \multicolumn{2}{@{}l}{$(x:o\to o)$}&&\\
  &\vline &\vline & & & & \multicolumn{2}{@{}l}{$=\langle p_4(p_6\linear p_7),\{x:p_4(p_6\linear p_7)\},
  \emptyset,\emptyset\rangle $}&&\\
  &\vline &\vline & & & \Sint & \multicolumn{2}{@{}l}{$(y:o)$}&&\\
  &\vline &\vline & & & & \multicolumn{2}{@{}l}{$=\langle p_8, \{y:p_8\}, \emptyset,\emptyset\rangle $}&&\\
  &\vline &\vline & & & \multicolumn{3}{@{}l}{\Unif$(p_6,p_8)=\{p_6=p_8\}$}&&\\
  &\vline &\vline & & & \multicolumn{3}{@{}l}{$=\langle p_7,\{x:p_6\linear p_7, y:p_6\}, \emptyset,
  \emptyset\rangle $}&&\\
  &\vline &\vline & & \multicolumn{4}{@{}l}{\Boxing$((x\ y),\{x:p_6\linear p_7, y:p_6\},p_7,
  \emptyset,\emptyset)$}&&\\
  &\vline &\vline & & & \multicolumn{3}{@{}l}{$=\langle\{x:b_1(p_6\linear p_7),
  y:b_1+p_6\},b_1+p_7, \emptyset\rangle $}&&\\ 
  &\vline &\vline & & \multicolumn{4}{@{}l}{\Unif$(b_1+p_7, p_2) = \{b_1+p_7-p_2 =
    0\}$}&&\\
  &\vline &\vline & & \multicolumn{4}{@{}l}{$cpts=\left\{\left(b_1+p_7-p_2=0,
  \left\{\begin{array}{l}
      x:b_1(p_6\linear p_7),\\
      y:b_1+p_6
    \end{array}\right\}\right)\right\}$}&&\\
  &\vline &\vline & & \multicolumn{4}{@{}l}{$=\left\langle
        p_3,\left\{\begin{array}{l} 
            x:p_2\linear p_3,\\
            x:b_1(p_6\linear p_7),\\
            y:b_1+p_6
          \end{array}\right\},\{b_1+p_7-p_2=0\},cpts\right\rangle$}&&\\ 
    \cline{4-9}
  &\vline & & \Boxing & \multicolumn{4}{@{}l}{$\left((x\ (x\ y)),\left\{\begin{array}{l}
        x:p_2\linear p_3,\\
        x:b_1(p_6\linear p_7),\\
        y:b_1+p_6
      \end{array}\right\}, p_3, cpts, \{b_1+p_7-p_2=0\}\right)$} &
      \multicolumn{2}{c|}{} \\
  &\vline & & & \Bang & \multicolumn{3}{@{}l}{$\left(\left\{\begin{array}{l}
        x:p_2\linear p_3,\\
        x:b_1(p_6\linear p_7),\\
        y:b_1+p_6
      \end{array}\right\}, p_3, cpts, \{b_1+p_7-p_2=0\}\right)$} &
      \multicolumn{2}{c|}{} 
\end{xtabular}

\begin{xtabular}{r@{}r@{}r@{}r@{}r@{}r@{}r@{}l|l|l|l|l|}
  &\vline & & & & \multicolumn{3}{@{}l}{$=\left\langle \left\{\begin{array}{l}
        x:b_2(p_2\linear p_3),\\
        x:b_1(p_6\linear p_7),\\
        y:b_1+p_6
      \end{array}\right\}, b_2+p_3,\{b_1+p_7-p_2-b_2=0\}\right\rangle$} & \multicolumn{2}{c|}{}\\
  &\vline & & & \multicolumn{4}{@{}l}{$=\left\langle \left\{\begin{array}{l}
        x:b_3+b_2(p_2\linear p_3),\\
        x:b_3+b_1(p_6\linear p_7),\\
        y:b_3+b_1+p_6
      \end{array}\right\},
        b_3+b_2+p_3,\{b_1+p_7-p_2-b_2=0\}\right\rangle$} &
        \multicolumn{2}{c|}{}\\
      &\vline & & \multicolumn{5}{@{}l}{$\mathcal{C}(b_3+b_2(p_2\linear p_3),b_3+b_1(p_6\linear p_7))
  =\left\{\begin{array}{l}
      b_3+b_2\ge 1\\
      b_3+b_2 = b_3+b_1\\
      p_2 = p_6\\
      p_3 = p_7
    \end{array}\right.=\left\{\begin{array}{l}
      b_3+b_2\ge 1\\
      b_2 = b_1\\
      p_2 = p_6\\
      p_3 = p_7
    \end{array}\right.$} & \multicolumn{2}{c|}{}\\
  &\vline & & \multicolumn{5}{@{}l}{$=\left\langle \begin{array}{ll}
        b_3+b_2(p_2\linear p_3)\linear b_3+b_2+p_3,&
      \{y:b_3+b_1+p_6\},\\
      \left\{\begin{array}{l} 
          b_1+p_7-p_2-b_2=0\\
          b_3+b_2\ge 1\\
          b_2 = b_1\\
          p_2 = p_6\\
          p_3 = p_7
        \end{array}\right.,&\emptyset\end{array} \right \rangle$} & \multicolumn{2}{c|}{}\\
&\vline & & \multicolumn{5}{@{}l}{$=\langle b_3+b_1(p_2\linear p_2)\linear b_3+b_1+p_2,
  \{y:b_3+b_1+p_2\}, \{b_3+b_1\ge 1\}, \emptyset \rangle$} &
\multicolumn{2}{c|}{}\\
\cline{2-10}
& & \Boxing & \multicolumn{5}{@{}l}{$(\lam x.(x\ (x\ y)), \{y:b_3+b_1+p_2\},
  b_3+b_1(p_2\linear p_2)\linear  b_3+b_1+p_2, \emptyset, \{b_3+b_1\ge
  1\})$}\\ 
& & & \multicolumn{5}{@{}l}{$=\langle \{y:b_2+b_3+b_1+p_2\}, b_2(b_3+b_1(p_2\linear
  p_2)\linear b_3+b_1+p_2),\{b_3+b_1\ge 1\}\rangle $}\\ 
& & \Unif & \multicolumn{5}{@{}l}{$(b_4((p_5\linear p_5)\linear p_5, b_2(b_3+b_1(p_2\linear
  p_2)\linear b_3+b_1+p_2))$}\\
& & & \multicolumn{5}{@{}l}{$=\left\{\begin{array}{l}
      b_4 = b_2\\
      p_5 = p_2\\
      p_5 = b_3+b_1+p_2
    \end{array}\right.=\left\{\begin{array}{l}
      b_4 = b_2\\
      p_5 = p_2\\
      b_3+b_1 = 0
    \end{array}\right.$}
\end{xtabular}

  Notice that the last constraint $b_3+b_1=0$ is incompatible with the
  previous $b_3+b_1\ge 1$ hence the set of solutions is empty.

\end{document}